\documentclass[%
 reprint,
 superscriptaddress,
 amsmath,amssymb,
 aps,
]{revtex4-2}

\usepackage{graphicx}
\usepackage{caption}
\usepackage{dcolumn}
\usepackage{bm}
\usepackage{gensymb}

\usepackage{titletoc}%!needs to be loaded before hyperref!
\usepackage{hyperref}
\usepackage[utf8]{inputenc}

\hypersetup{
    colorlinks,
    citecolor=black,
    filecolor=black,
    linkcolor=black,
    urlcolor=black
}

\makeatletter
\def\maketitle{
\@author@finish
\title@column\titleblock@produce
\suppressfloats[t]}
\makeatother

\preprint{APS/123-QED}
\begin{document}

\title{Quantized critical supercurrent in  SrTiO$_3$-based  quantum point contacts}

\author{Evgeny Mikheev}
\affiliation{Department of Physics, Stanford University, Stanford, CA, 94305, USA}
\affiliation{Stanford Institute for Materials and Energy Sciences, SLAC National Accelerator Laboratory, Menlo Park, California 94025, USA}
\author{Ilan T. Rosen}
\affiliation{Department of Applied Physics, Stanford University, Stanford, CA, 94305, USA}
\affiliation{Stanford Institute for Materials and Energy Sciences, SLAC National Accelerator Laboratory, Menlo Park, California 94025, USA}
\author{David Goldhaber-Gordon}
\affiliation{Department of Physics, Stanford University, Stanford, CA, 94305, USA}
\affiliation{Stanford Institute for Materials and Energy Sciences, SLAC National Accelerator Laboratory, Menlo Park, California 94025, USA}
% \date{\today}

\captionsetup[figure]{labelfont={bf},labelformat={default},labelsep=period,name={Fig.},justification=justified}

\begin{abstract}

Superconductivity in SrTiO$_3$ occurs at remarkably low carrier densities and therefore, unlike conventional superconductors, can be controlled by electrostatic gates. Here we demonstrate nanoscale weak links connecting superconducting leads, all within a single material, SrTiO$_3$. Ionic liquid gating accumulates carriers in the leads, and local electrostatic gates are tuned to open the weak link. These devices behave as superconducting quantum point contacts with a quantized  critical supercurrent. This is a milestone towards establishing SrTiO$_3$ as a single-material platform for mesoscopic superconducting transport experiments, that also intrinsically contains the necessary ingredients to engineer topological superconductivity.
\end{abstract}
\maketitle

\section*{Introduction}
Conductance quantization in ballistic quantum point contacts (QPC) is a striking example of departure from the classical Drude picture of electrical conductivity set by the rate of charge carrier scattering \cite{Buttiker90}. When a constriction between two electron reservoirs is sufficiently narrow and disorder-free, its conductance becomes quantized according to the number of occupied modes: discrete transverse momenta allowed within the constriction's confinement potential. Each mode contributes a conductance quantum $\delta G = 2e^2/h$ (spin-degenerate case), a value that does not depend on the exact geometry of the device.

A related phenomenon is expected to arise in a constriction between two superconducting reservoirs \cite{Beenakker91, Furusaki92}, i.e. a superconducting quantum point contact (SQPC). Again, the transverse momentum spectrum becomes discretized under the constriction confinement potential. The supercurrent carried by each mode is determined by the Andreev bound state (ABS) spectrum, which is typically a function of constriction geometry. SQPCs are thus characterized by quantized critical supercurrent $I_C$ with a non-universal step height $\delta I_C$. However, in the limit of a short junction length, only one ABS per ballistic mode remains, and the current carried by each mode can reach a maximum value $\delta I_C = e\Delta/\hbar$. This ideal step height is again geometry-independent and scales only with the superconducting gap $\Delta$.

The widespread route for fabricating gate-tunable superconducting weak links has been to combine two optimal components in a hybrid system: a clean semiconductor (typically a III-V semiconductor or Ge) and metallic superconducting leads  (for example Nb, Al). Such hybrid systems have been successfully used to demonstrate quantized critical supercurrent, but with quantization step heights far below $e\Delta/\hbar$ \cite{Takayanagi95,Bauch05,Xiang06,Abay13,Irie14,Hendrickx19}. The two major challenges for reaching the universal limit for quantized supercurrent are the geometric requirement that the distance between superconducting leads be much less than the superconducting coherence length $\xi$ and the need for near perfect semiconductor/superconductor contact transparency \cite{Furusaki92}. Achieving the latter in hybrid semiconductor-superconductor systems has been a major materials science challenge that has required deployment of in-situ heteroepitaxial growth techniques \cite{Kjaergaard17}.

An alternate route taken in this work is to form both leads and constriction in a single electrostatically tunable superconducting material, such as SrTiO$_3$ (STO). Working within a single material platform  is attractive for fabricating SQPCs, as the SN boundary can be purely electronic (no structural discontinuity) and thus potentially highly transparent.

One of STO's remarkable aspects is superconductivity in the extremely dilute charge carrier density limit \cite{Lin13,Gastiasoro20}. In 2D electron systems (2DESs) at the surface of STO, such as LaAlO$_3$/SrTiO$_3$ (LAO/STO), LaTiO$_3$/SrTiO$_3$ and ionic liquid-gated STO, superconductivity occurs in the range of 0.01 electrons per unit cell \cite{Bell09, Joshua12}. Consequently, one can electrostatically control the transition between superconductor, normal metal, and insulator in this material. On the macroscopic scale such control is well established using back-gating through the STO substrate, top-gating through a dielectric layer, and ionic liquid gating \cite{Caviglia08,Bell09, Joshua12, Gallagher15, Chen18, Christensen19}.

\begin{figure*}
\centering
\includegraphics[width=3.155in]{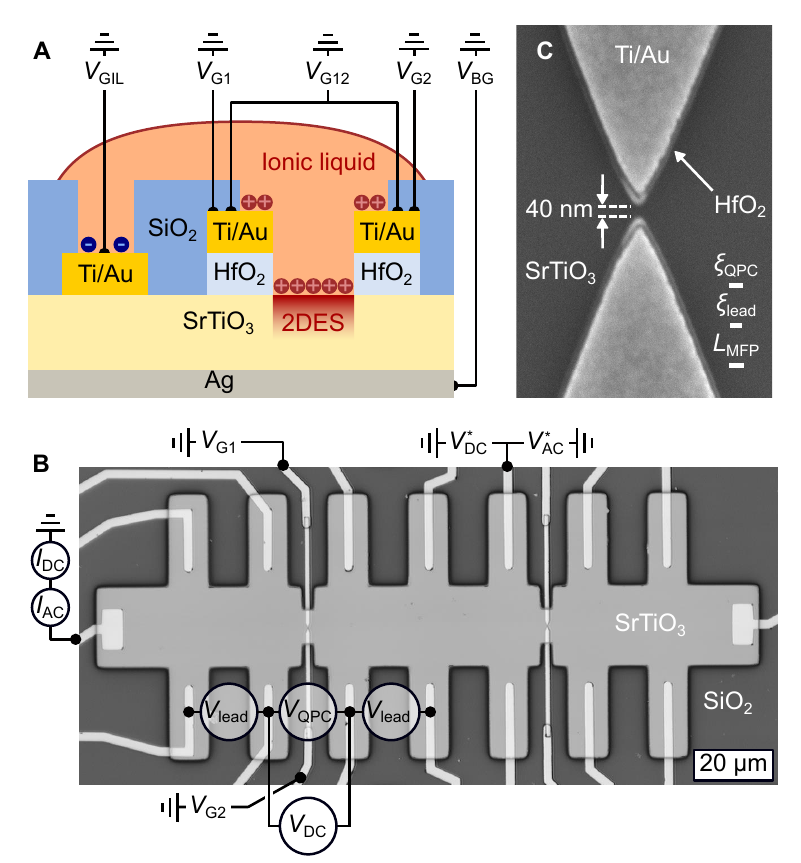}
\includegraphics[width=3.845in]{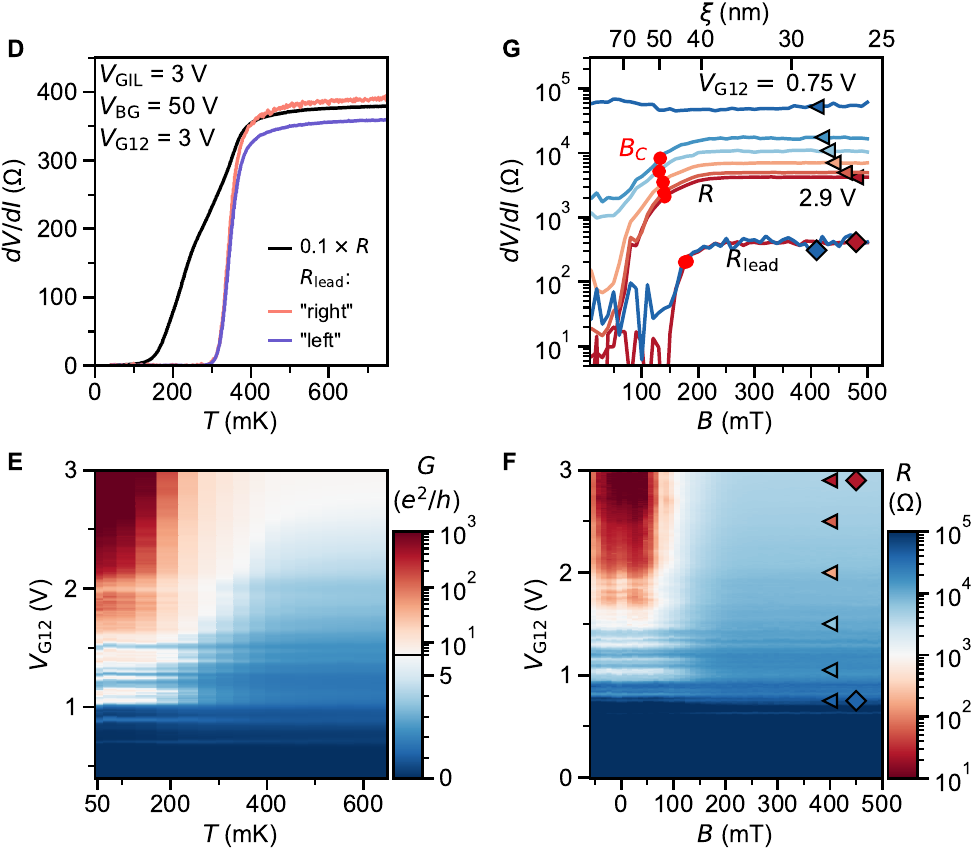}
\caption{\label{f1} \textbf{Electrostatically defined constriction in superconducting SrTiO$_3$.} (\textbf{A}) Schematic cross-section of the device, and illustration of the gate voltage definitions. (\textbf{B}) Confocal laser microscope image of the Hall bar region of the device, and illustration of the measurement scheme. (\textbf{C}) Scanning electron microscope image of the constriction region on a reference device. (\textbf{D}) Superconducting transition in the constriction and lead resistance. ``Right" and ``left" refer to measurement of $V_\text{lead}$ on both sides of the constriction. (\textbf{E}) Constriction conductance map with temperature and split gate voltage. (\textbf{F}) Constriction conductance map with magnetic field and local gate voltage. Symbols in (\textbf{F}) indicate the selected gate voltage values for which line cuts in field are shown in (\textbf{G}). Lead resistance at extremes of $V_\text{G12}$ is also shown in (\textbf{G}) to illustrate independence of local gate voltage. The top axis shows the mapping from critical field $B_C$ (red circles) to the coherence length $\xi$ . The estimated $\xi$ is shown in (\textbf{C}) for comparison with device dimensions, along with the mean free path from Hall measurements in the leads (see supplementary section S4). In (\textbf{D}-\textbf{G}), $V_\text{GIL} = $ 3~V,  $V_\text{BG} = $ 50~V.}
\end{figure*}

More recently, several approaches have emerged for nanoscale patterning of conduction in LAO/STO, leading to demonstration of quantization effects in normal state, but so far not in superconducting transport. Realization of a conventional split-gate QPC geometry in LAO/STO is challenging, as it involves depleting and/or accumulating charge densities of at least $\approx 10^{13}$~cm$^{-2}$, close to the limit of conventional dielectrics. Spatial inhomogeneity and relatively short mean free paths in such 2DESs are another challenge, leading patterned constrictions to often be dominated by tunneling through accidental quantum dots \cite{Maniv16,Prawiroatmodjo17,Thierschmann18}. A QPC with normal state but not superconducting conductance quantization has recently been demonstrated in underdoped, non-superconducting LAO/STO \cite{Jouan20}. In  \cite{Thierschmann18}, a constriction defined by split gates with normal state conductance about half of a single spin-degenerate ballistic mode was estimated to have corresponding partially transmitting single-mode supercurrent, though it did not show direct effects of quantization. A different technique is to write conductive channels on LAO/STO with voltage-biased AFM tips. This method enabled demonstration of quantum wires and dots coupled by tunnel barriers to superconducting leads, with quantized normal-state transport and indirect signatures of electron pairing \cite{Cheng15,Cheng16,Annadi18,Briggeman20} but not superconductivity.

In this work, we demonstrate quantized supercurrent in quantum point contacts in a split-gate geometry, based on ionic liquid gated SrTiO$_3$. We demonstrate a quantized critical current, with tuning from zero to three ballistic modes. Step height per mode $\delta I_C$ is only $3-5$x smaller than the canonical value $e\Delta / \hbar$, as close to ideal as achieved in any hybrid system \cite{Xiang06}. The fabrication process of our devices is enabled by the fine patterning of local electrostatic gates, using lift-off of metal and atomic layer deposited Hafnia (HfO$_2$) with feature size close to 40~nm. This is distinct from the approaches taken in previous works on LAO/STO weak links \cite{Goswami15,Monteiro17, Stornaiuolo17, Prawiroatmodjo17,Thierschmann18,Jouan20}. Notably, we avoid an epitaxial growth step at high temperature, which complicates the workflow for patterning and potentially introduces disorder (see e.g. \cite{balakrishnan19,Schneider19}). We thus consider this fabrication technique an attractive alternative for further development of STO as a platform for mesoscale superconducting devices.

\section*{Results}

Our devices are 20~$\mu$m wide Hall bars covered by ionic liquid, which is polarized to accumulate a 2D carrier density at any exposed STO surface. The coarse contours of the Hall bar are defined by patterning an insulating SiO$_2$ layer which separates the surface of undoped STO from the ionic liquid (Fig.~\ref{f1}A,B); underneath the SiO$_2$, the STO surface remains insulating, while the carrier density in the Hall bar region is tuned into the superconducting regime. Split gates with thin, self-aligned HfO$_2$ dielectrics define 40~nm wide constrictions (Fig.~\ref{f1}C) between neighboring superconducting reservoirs. The design includes 5 or 6 ohmic contacts on each side of the split gates (Fig.~\ref{f1}B) to enable four-terminal measurements of both the constriction and the adjacent superconducting leads.

The carrier density profile is electrostatically defined by voltages on four gates, as illustrated in Fig.~\ref{f1}A: a large coplanar gate that controls the polarization of the ionic liquid ($V_\text{GIL}$), a back gate ($V_\text{BG}$) and two split gates ($V_\text{G1}$ and $V_\text{G2}$, denoted as $V_\text{G12}$ for the case $V_\text{G1} = V_\text{G2}$). $V_\text{GIL}$ is set when the device is near room temperature, and maintained as the sample is cooled below the freezing temperature of the ionic liquid (220~K).  $V_\text{GIL}$ is used to polarize a drop of ionic liquid that covers both the coplanar gate electrode and the device. At lower temperatures the polarization of the ionic liquid is frozen in. $V_\text{GIL}$ is the primary control knob for the carrier density in the  leads, which can be tuned from $\approx 5 \times 10^{12}$ to $10^{14}$~cm$^{-2}$ \cite{Ueno08, Lee11}.
The superconducting transition temperature as a function of density has a maximum near $3 \times 10^{13}$~cm$^{-2}$ (see supplementary section S3). The main results presented in this paper will focus on this nearly optimally doped state, obtained by cooling the device under $V_\text{GIL} = $ +3~V. For additional data on the second constriction on the right side of the Hall bar in Fig.~\ref{f1}B,  different devices and cool-downs with carrier density tuned across a larger range, see supplementary sections S2-S6.

\begin{figure}[!b]
\centering
\includegraphics[width=3.5in]{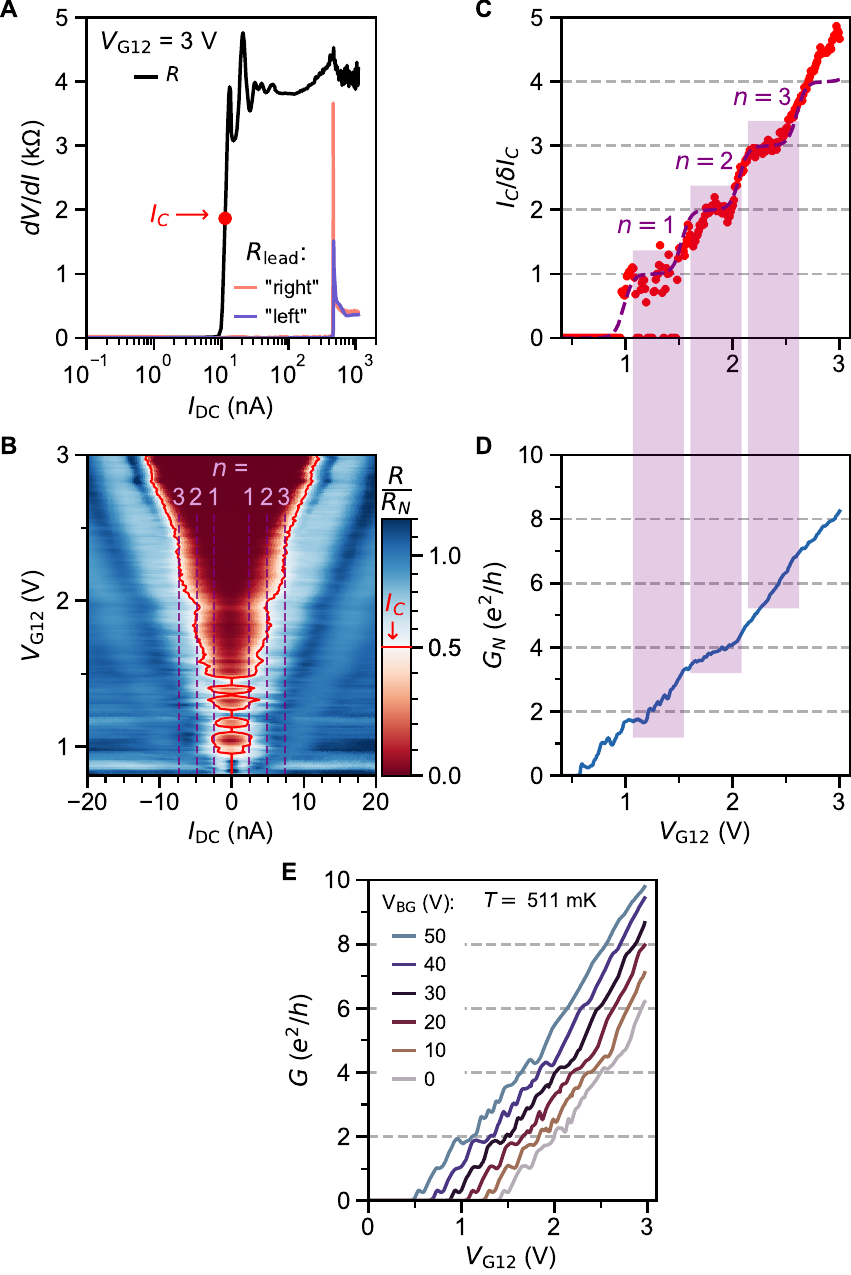}
\caption{\label{f2}\textbf{Critical current quantization.} (\textbf{A}) DC current dependence of constriction and lead resistances at $V_\text{G12} =$ 3V. (\textbf{B}) Constriction resistance, normalized to normal state resistance at $V_\text{DC} = $ 100 $\mu$V. The solid red line indicates the critical current $I_C$. The dashed lines indicate 1, 2 and 3 integer multiples of $\delta I_C = $ 2.48 nA. (\textbf{C}) $V_{G12}$ dependence of $I_C$ normalized to $\delta I_C$ and (\textbf{D}) normal state conductance $G_N$ at $V_\text{DC} = $ 100 $\mu$V, with a series resistance of 800~$\Omega$ subtracted from the raw data. The shaded connection between (\textbf{C}) and (\textbf{D}) emphasizes the numerical correspondence in the observed number of ballistic modes $n$. The dashed line in (\textbf{C}) is a fit to the saddle potential QPC model (see supplementary section S1). (\textbf{E}) Split- and back-gate voltage dependence of zero-bias conductance above $T_C$. $G$ has been corrected for a variable series resistance gradually increasing from 1.15 to 2.1 k$\Omega$. Short plateaus can be seen at integer multiples of $2e^2/h$ ($n=$ 1, 2, and hints at higher multiples). Unintentional coulomb blockade levels can be seen near 0.2$e^2/h$, $e^2/h$ and 2.5$e^2/h$.}
\end{figure}

The voltage $V_\text{BG}$ on a back gate contacting the bottom of the SrTiO$_3$ crystal provides additional global tuning of the 2DES at base temperature, primarily by modulating the depth of the 2DES. For most experiments on this device, we set $V_\text{BG} = $ +50~V to pull the electron density farther away from surface disorder (see \cite{Chen16} and supplementary section S4).

Fig.~\ref{f1}D shows the superconducting transition $T_C$ measured by sourcing a small AC excitation through a constriction at $V_\text{G12} = $ +3~V and $V_\text{BG} = $ +50~V. In the following, constriction resistance and conductance will be denoted as $R=dV_\text{QPC}/dI_\text{AC}$ and $G=1/R$, and the resistances of the leads as $R_\text{lead}=dV_\text{lead}/dI_\text{AC}$ (see Fig.~\ref{f1}B  and the Methods section for more details). On both sides of the constriction, $R_\text{lead}$ shows a sharp transition near 350~mK. This is near the optimal value for 2D SrTiO$_3$ \cite{Joshua12,Chen18}. The measured Hall density of $3.05 \times 10^{13}$~cm$^{-2}$ and the slight increase of $T_C$ by 20 mK upon removing the back-gate voltage suggest that this device state is slightly on the overdoped side of the superconducting dome (see  supplementary section S4).

The constriction resistance $R$ also starts decreasing near the lead $T_C$, but its transition to zero resistance (within accuracy of our measurement) is significantly broader than that of the leads. Decreasing $V_\text{G12}$ suppresses both the zero resistance state and the normal state conductance, and eventually pinches off the weak link (Fig.~\ref{f1}E,F). At base temperature, superconductivity can also be suppressed by a perpendicular magnetic field (Fig. 1f). Using $\xi^2=\Phi_0/(2\pi B_\text{C})$ \cite{Kim12}, with $\Phi_0=h/2e$ being the flux quantum, the critical field $B_{C} =$ 130-140~mT in the constriction yields an estimated coherence length $\xi = $ 50~nm (43~nm in the leads). This estimate is consistent with the dirty-limit BCS superconductor picture \cite{Bert12, Collignon17}, in which the coherence length is set by the mean free path $L_\text{MFP}$. From Hall measurements on the leads, we extract a Hall mobility $\mu =$ 600~cm$^2$/Vs and $L_\text{MFP} = $ 55~nm.

The shortness of these length scales illustrates the challenge of fabricating QPCs and SQPCs in SrTiO$_3$ (see Fig.\ref{f1}C). Observing ballistic transport requires junction length $L<L_\text{MFP}$. Achieving a single-ABS junction with critical current quantization also requires short junction length: $L < \xi$. Though the junction length is not well defined in a split gate geometry, we fabricated the gates with very narrow lateral spacing (40~nm) and sharp tips to strive for the ballistic (or quasi-ballistic) regime.

The ballistic nature of the SQPC is most apparent in differential resistance at finite DC current. Filling of states in the constriction with $V_\text{G12}$ results in a staircase shape of the critical supercurrent $I_C(V_\text{G12})$ (Fig.~\ref{f2}). Adopting a  definition of $I_C$ as the current at which the normal state resistance is halved, plateaus at both positive and negative integer multiples of $\delta I_C = $ 2.48 nA are seen in the $V_\text{G12}-I_\text{DC}$ map of constriction resistance normalized to its normal state value (Fig.~\ref{f2}B). 

In the ballistic SQPC picture, $I_C/\delta I_C$ corresponds to $n$, the number of ballistic modes below the Fermi energy in the constriction (Fig.~\ref{f2}C).The first mode plateau is intermittent as a function of gate voltage due to resonant transmission through the weak link, correlated with the charging levels of an accidental coulomb blockade observed near pinch-off at low $V_{G12}$ (see supplementary section S7), whereas the second and third plateaus are more stable. An alternative way to estimate the number of modes is from normal state conductance $G_N$, where each fully transmitting spin-degenerate mode is expected contribute a conductance $\delta G = 2e^2/h$. The number of modes inferred by dividing $G_N$ by this increment matches that extracted from the sequence of steps in supercurrent. We also see hints of plateaus in normal state conductance near $n = $ 1 and 2 (Fig.~\ref{f2}D). Features suggestive of normal state conductance quantization are more clearly apparent above $T_C$  (Fig.~\ref{f2}E), where one does not need to apply a DC bias to suppress the supercurrent, and disorder-induced fluctuations are reduced. The plateau structure persists as a function of back gate voltage, as detailed further in supplementary section S5.

\begin{figure}
\centering
\includegraphics[width=3.5in]{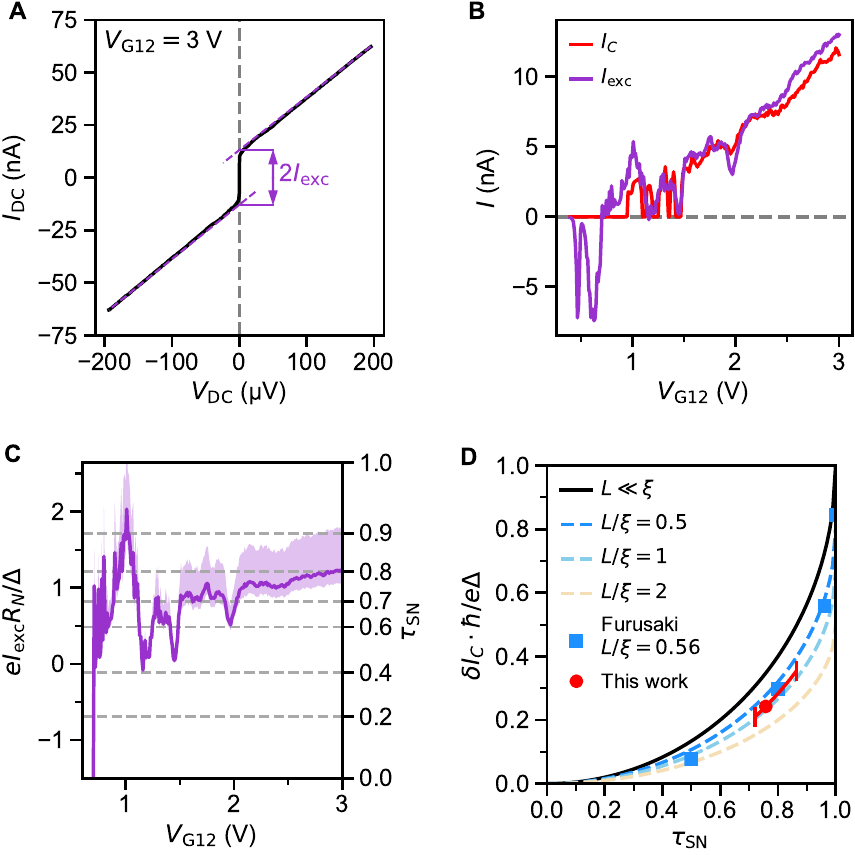}
\caption{\label{f3}\textbf{SN transparency and junction length.} (\textbf{A}) The DC current-voltage curve of the constriction at $V_\text{G12}=$ 3~V and the definition of the excess current $I_\text{exc}$. (\textbf{B}) Split gate voltage dependence of the excess and critical currents. (\textbf{C}) $eI_\text{exc}R_N/\Delta$, the input quantity of the SNS model in ref \cite{Octavio83,Flensberg88}, and its mapping onto SN boundary transparency $\tau_\text{SN}$. (\textbf{D}) $\delta I_C$ suppression by finite transparency, and finite junction length. Comparison to the ballistic short-limit model \cite{Beenakker91} (solid black line), full calculation at $L/\xi=$ 0.56 \cite{Furusaki92} (blue squares,) and with the approximate correction for arbitrary $L/\xi$ from \cite{Bagwell92} (dashed lines). The shaded region reflects in (\textbf{C}) and error bars in (\textbf{D}) reflect the uncertainty on the gap, see supplementary section S8.}
\end{figure}

Ideally, the magnitude of steps in $I_C$ through a constriction should scale only with the superconducting gap as
\begin{equation}
\delta I_C=\frac{e\Delta}{\hbar}.
\end{equation}
This scaling is expected to hold for a short junction ($L<<\xi$) with perfectly transparent SN contacts \cite{Beenakker91, Furusaki92}.

For most experimental realizations of SQPC's in hybrid metal superconductor/semiconductor devices, neither of these requirements is fully satisfied, and $\delta I_C$ is generally suppressed by at least an order of magnitude \cite{Takayanagi95,Bauch05,Abay13,Irie14,Hendrickx19}. One work on Si/Ge nanowires with Nb contacts reported suppression by only a factor of 2.9 \cite{Xiang06}. In our case, data in fig.~\ref{f2} suggests a comparable factor of 3-5. The uncertainty comes from the choice of method to extract $\Delta$ (see supplementary section S8): from $T_C$ of the constriction [$\delta I_C/(e\Delta/\hbar) =$ 2.9], from $T_C$ of the leads [$\delta I_C/(e\Delta/\hbar) =$ 4.1], or from the temperature dependence of the excess current [$\delta I_C/(e\Delta/\hbar) =$ 4.8].

Analysis of the excess current $I_\text{exc}$ allows separating the role of imperfect SN contact transparency $\tau_\text{SN}$ from that of finite junction length. We define $I_\text{exc}$ as the zero-bias intercept of the normal-state resistance extrapolated from high $V_\text{DC}$ (Fig.~\ref{f3}A). Its evolution with $V_\text{G12}$ approximately tracks that of $I_{C}$. The quantity $eI_\text{exc}R_N/\Delta$ can be non-linearly mapped onto $\tau_\text{SN}$ following the treatment of Andreev reflections in an SNS junction in \cite{Octavio83, Flensberg88}. Over the gate voltage range with a well defined and quantized supercurrent ($1.5 < V_\text{G12} <  2.5$), we thereby extract $\tau_\text{SN} = 0.75^{+0.12}_{-0.08}$.

In the short junction limit $L\ll\xi$ \cite{Beenakker91}, we can predict the suppression of $\delta I_{C}$ as a function of $\tau_\text{SN}$ (Fig.~\ref{f3}D). The experimentally measured $\delta I_C$ is only slightly below the theoretical curve, and its full suppression can be accounted for by multiplying it by an additional factor $\alpha = $ 0.7. This additional suppression can be explained by considering the finite length of the junction. An approximate theoretical description obtained in \cite{Bagwell92} is $\alpha=1/(1+L/2\xi)$, which is in good agreement with calculations for the case in \cite{Furusaki92}, where $L=0.56\xi$. In this work, assuming $\alpha=0.7$ yields $L =  0.85 \xi = 42$~nm, which is close to the 40~nm lithographic width of our QPC. 

\section*{Discussion}

So transparency is likely to be the main driver for the reduction in $\delta I_{C}$ from its ideal value, despite being competitive with the hybrid III-V/superconductor systems, where $\tau_\text{SN}$ is typically estimated below 0.85 \cite{Abay13, Irie14,Li16, Hendrickx19} except for pristine epitaxial interfaces \cite{Kjaergaard17}. An advantage of our single-material system is that the SN contact interface is electrostatically defined and presumably does not have a structural discontinuity. In our present realization, transparency is likely limited by the smooth gate-induced density variation which in turn entails a gradually varying order parameter. We anticipate that $\tau_\text{SN}$ can be further improved by manipulating the SN boundary with additional local gates near the weak link.

Furthermore, we anticipate improvements by increasing the mean free path. In ionic liquid-gated STO and LAO/STO, $L_\text{MFP}$ is typically less than 100~nm. However, improvements to $\mu > 10^4$ cm$^2$/Vs and $L_\text{MFP} > $ 1~$\mu$m have been demonstrated by separating the ionic liquid from the channel by an ultrathin spacer layer \cite{Gallagher15}, band engineering with spacer layers in LAO/STO \cite{Chen15}, or forming the channel from high quality MBE-grown STO in the 3D case \cite{Son10}. The fabrication route used in this work is relatively simple -- based on commercially available STO crystals, avoiding epitaxial growth steps -- so complex patterning or design refinements could be added without rendering it unwieldy.

% \section*{Conclusion}

Using ionic liquid gated STO as a platform, we have realized SQPCs with quantized critical supercurrent, tunable between zero and three ballistic modes by split gates. This is a first realization of a gate-tunable SQPC in a single material system, enabling highly transparent SN contacts without structural discontinuity at the boundary. This work establishes spatially-patterned screening of ionic liquid from an STO surface as a promising alternative to existing methods for nanoscale patterning of conduction and superconductivity in STO: patterning LAO/STO with pre-growth templates \cite{Maniv16,Monteiro17, Stornaiuolo17,Prawiroatmodjo17}, electrostatic depletion by patterned gates \cite{Goswami15,Thierschmann18,Jouan20}, or conductive channel writing by voltage biased AFM tips \cite{Cheng15,Cheng16,Annadi18,Briggeman20,Boselli20}. Our method appears particularly suited for realizations of ballistic superconducting transport, which requires maintaining a high carrier density within nanopatterned constrictions. Naturally-occurring depletion near the edges of an STO-based conducting channel \cite{Persky20,Boselli20} can be counteracted with local gates as we have shown.

Our approach may also be especially attractive for exploring topological superconductivity in several contexts. Combining ballistic transport with superconductivity, strong spin-orbit coupling, and tunable dimensionality offers hope for engineering extrinsic topological superconductivity in one-dimensional nanostructures \cite{Fidkowski13,Mazziotti18,Perroni19}. Even an unpatterned SrTiO$_3$ 2DES may host intrinsic topological superconductivity in certain conditions due to interplay between its multi-orbital band structure, spin-orbit coupling, and ferroelectricity \cite{Scheurer15,Kanasugi18,Kanasugi19}. A ballistic point contact similar to the SQPC demonstrated here could serve as the tunnel probe central to many detection schemes for the resulting Majorana bound states \cite{Wimmer11,Prada12,Zhang19}. The single-mode ballistic Josephson junction regime demonstrated here is also a requisite ingredient of theoretical proposals for realizing topological Andreev bound state spectra in multi-terminal junctions \cite{Riwar16,Xie17}. Finally, this work is an important step toward realizing controlled negative-$U$ quantum dots \cite{Cheng15,Prawiroatmodjo17} in the classic geometry of an ``island" coupled to two QPCs \cite{Hanson07}.

\section*{Materials and Methods}

Fabrication is based on commercial (001)-oriented SrTiO$_3$ single crystal substrates, purchased from MTI. To obtain a Ti-terminated surface with terrace step morphology, these substrates were soaked in heated deionized water for 20 minutes and annealed at 1000  \degree C for 2 hours in flowing Ar and O$_2$ in a tube furnace.

All subsequent patterning was performed with lift-off processes using e-beam patterned PMMA 950K, 4\% in anisole for the first step, 8\% for all subsequent steps. The first step is the local split gate pattern, written on a 100 kV e-beam write system. Atomic layer deposition was used to deposit 15~nm HfO$_2$ (100 cycles of Hf precursor and water.) The deposition stage temperature was 85  \degree C. We note the importance of loading the sample and starting the deposition quickly to avoid PMMA pattern reflow. The 5~nm Ti / 50~nm Au gate contact was then deposited by e-beam evaporation. Lift-off of both HfO$_2$ and Ti/Au layers was then performed by soaking in heated NMP, followed by ultrasonication in acetone.

The remaining patterning was performed with a 30 kV e-beam write system. The second step is the gate contact, using lift-off of 40~nm Ti / 100~nm Au in acetone. The third step is the ohmic contact deposition. It requires exposing the pattern to Ar$^+$ ion milling prior to e-beam evaporation of 10~nm Ti / 80~nm Au, followed by lift-off in acetone. The fourth patterning step is the mesa insulation, deposited by magnetron sputtering 70~nm of SiO$_2$, followed by lift-off in acetone. The measured devices were imaged with a conventional optical microscope and with a Keyence VK-X confocal laser microscope. Scanning electron microscope imaging was performed on reference patterns written on the same chips.

Finished devices were annealed for 20 minutes at 150 \degree C in air. The back gate contact to a gold pad on an alumina ceramic chip carrier was made with silver paste. Immediately after depositing a drop of ionic liquid Diethylmethyl(2-methoxyethyl)ammonium bis(trifluoromethylsulfonyl)imide (DEME-TFSI) to cover both the device and the surrounding side gate, the samples were loaded into the dilution refrigerator system, then vacuum pumped overnight to minimize contamination of the ionic liquid by water from exposure to air. 

The ionic liquid gate voltage $V_\text{GIL}$ was slowly ramped up to desired value at room temperature, followed by several minutes of stabilization and then rapid cooling the measurement probe below the freezing point of DEME-TFSI (220~K).

Typical measured resistance per successful ohmic contact was 3-10~k$\Omega$, which includes a 2-3~k$\Omega$ contribution from the measurement lines and built-in RF filters in the probe. Measurements were performed by voltage sourcing nominal AC and DC excitations ($V_\text{AC}^{*}$ and $V_\text{DC}^{*}$) through an adder circuit and measuring the drained current. $V_\text{AC}$ and $V_\text{DC}$ refer to the measured AC and DC components of $V_\text{QPC}$, the voltage drop across the weak link.

% \defbibheading{bibliography}[\refname]{\section*{#1}}
% \printbibliography

\renewcommand{\bibsection}{\section*{References}}
\bibliographystyle{apsrev4-1}
\bibliography{references.bib}

\section*{Acknowledgments}
We acknowledge Marc Kastner, Malcolm Beasley and Eli Fox for helpful discussions. We acknowledge Richard Tiberio and Michelle Rincon for help with device fabrication.

\textbf{Funding:} This work was supported by the Air Force Office of Scientific Research through grant no. FA9550-16-1-0126. E. M. was supported by the Nano- and Quantum Science and Engineering Postdoctoral Fellowship at Stanford University. I. T. R. was supported by the U.S. Department of Energy, Office of Science, Basic Energy Sciences, Materials Sciences and Engineering Division, under Contract DE-AC02-76SF00515. Measurement infrastructure was funded in part by the Gordon and Betty Moore Foundation’s EPiQS Initiative through grant GBMF3429 and grant GBMF9460. Part of this work was performed at the Stanford Nano Shared Facilities (SNSF)/Stanford Nanofabrication Facility (SNF), supported by the National Science Foundation under award ECCS-1542152.

\textbf{Author contributions:} E.M. and D.G.-G. designed the experiment. E.M. fabricated the devices. E.M. and I.R. performed the measurements. E.M. carried out data analysis. All authors discussed the results and wrote the manuscript.

\textbf{Competing interests:} The authors declare that they have no competing interests. 

\textbf{Data and materials availability}: All data needed to evaluate the conclusions in the paper are present in the paper and/or the Supplementary Materials. Additional data related to this paper may be requested from the authors.

\clearpage

\title{Supplementary material for ‘‘Quantized critical supercurrent in  SrTiO$_3$-based  quantum point contacts’’}

\maketitle

\clearpage

\onecolumngrid

% \begin{center}
% \textbf{\large Supplementary Information}
% \end{center}

\captionsetup[figure]{labelfont={bf},labelformat={default},labelsep=period,name={Fig.}}

\setcounter{figure}{0}
\renewcommand{\thefigure}{S\arabic{figure}}

\setcounter{page}{1}
\renewcommand{\thepage}{S\arabic{page}}
\setcounter{section}{0}
\renewcommand{\thesection}{S\arabic{section}}
\setcounter{equation}{0}
\renewcommand{\theequation}{S\arabic{equation}}

% \renwecommand{\cite}{\textsuperscript{\cite{#}}}
% Supplementary material for ‘‘Quantized critical supercurrent in  SrTiO$_3$-based  quantum point contacts’’

% \tableofcontents

\textbf{This PDF file includes:}

Supplementary Text

Figs. S1 to S40

\clearpage

\section{Theoretical framework}
\subsection{The nature of the SNS weak link}

\begin{figure*}[b]
\centering
\includegraphics[width=7in]{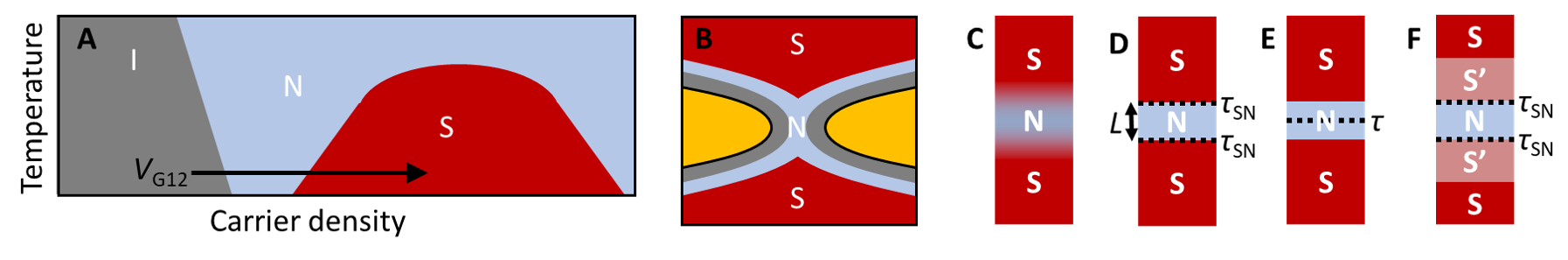}
\caption{\label{SM_sketch1} (\textbf{A}) Schematic phase diagram of SrTiO$_3$ as a function of doping. (\textbf{B}) Schematic top view of a normal state constriction with superconducting leads. (\textbf{C}-\textbf{F}): different 1D model representations of the constriction: (\textbf{C}) SNS with diffuse SN boundaries due to gradual change in carrier density, (\textbf{D}) idealized SNS with sharp SN boundaries with transparency $\tau_\text{SN}$, (\textbf{E}) same as (\textbf{D}) but with an alternate definition of junction transparency $\tau = \tau_\text{SN}^2$, (\textbf{F}) SS'NS'S constriction, where S' is a superconducting region with a reduced order parameter in comparison to the S region.}
\end{figure*}

SrTiO$_3$ can be described as a semiconducting superconductor: as a function of carrier density, its ground state evolves from an insulator to a normal metal to a superconductor. A schematic phase diagram is shown in Fig.~\ref{SM_sketch1}A, roughly summarizing a very large body of work on substitutionally-doped SrTiO$_3$, LaAlO$_3$/SrTiO$_3$, LaTiO$_3$/SrTiO$_3$, and ionic liquid-gated SrTiO$_3$, see e.g. \cite{Lin13,Caviglia08,Joshua12,Gallagher15,Christensen19}.

Fig.~\ref{SM_sketch1}B illustrates our understanding of how the split gates create a weak link between superconducting reservoirs by locally depleting the carrier density. At higher positive split gate voltage, an SNS junction is formed. At lower gate voltage, the depletion region extends further into the constriction, eventually pinching it off. In this simplistic picture, the underdoped side of the phase diagram is reproduced as a function of distance from the split gate. The resulting weak link is then likely to not have sharp SN boundaries, but instead gradual transitions from near-optimal $T_c$ to weak superconductivity and then to normal metal (Fig.~\ref{SM_sketch1}C).

A complete modelling of such a system is a difficult task, in particular due to the non-linear dielectric constant of SrTiO$_3$ \cite{Thierschmann18}, and complex interplay between microscopic pairing and macroscopic coherence in the underdoped regime \cite{Prawiroatmodjo16,Singh18,Chen18}. 

For simplicity, we choose to model our devices as SNS junctions, with normal region length $L$ and abrupt SN interfaces with an effective transparency $\tau_\text{SN}$ (Fig.~\ref{SM_sketch1}D). In some cases, a single transparency $\tau$ is defined for the entire junction (Fig.~\ref{SM_sketch1}E); we approximate the relation between the two as $\tau=\tau_\text{SN}^2$, assuming no scattering within the N region.

A potential added complexity that will need to be addressed in follow up work is whether an SS'NS'S description (Fig.~\ref{SM_sketch1}F) is more appropriate for such junctions than SNS. S' is either a superconducting region with the order parameter reduced by depletion, or a normal metal with a pairing gap induced by proximity effect  \cite{Golubov95}. In either case, the S' pairing scale becomes distinct from the S scale measured in the leads. In the case of a short S' region, both the S and S' scales are relevant for Josephson and tunneling transport \cite{Golubov95,Aminov96}.

\subsection{Critical current of a short junction with finite contact transparency}

Following \cite{Beenakker91,Furusaki99}, the simplest picture of a ballistic SQPC is given by a one-dimensional, ‘‘short-limit’’ SNS model. The pair potential is taken to be a step function: bulk-like in the S region, and zero in the N region. Imperfect junction transparency is modeled by introducing scattering in the N region from a $\delta$-function potential
\begin{equation}
U(x)=V_{B}\delta(x),
\end{equation}
which is traditionally renormalized into a dimensionless parameter $Z=mV_{B}/\hbar^2k_F$, with $k_F$ being the Fermi wavevector. The equivalent transmission probability of the N region is $\tau=1/(1+Z^2)$. In the short limit, where the length of the junction is much shorter than the superconducting coherence length ($L\ll\xi$), the phase dependence of the Andreev bound state (ABS) spectrum is given by \cite{Beenakker91}
\begin{equation}
E_B(\phi)=\Delta \sqrt{1-\tau \sin ^2 (\phi/2)},
\end{equation}
and the current-phase relationship for a single ballistic mode is
\begin{equation}
\label{eqSNSIc}
I_1(\phi) = \frac{e\Delta}{\hbar}\cdot
\frac{\sin(\phi)}{2}\cdot
\sqrt{\frac{\tau}{\cos^2(\phi/2)-1+\tau^{-1}}}\cdot
\tanh\left(\frac{E_B}{2k_B T}\right).
\end{equation}

\begin{figure*}[b]
\centering
\includegraphics[width=7in]{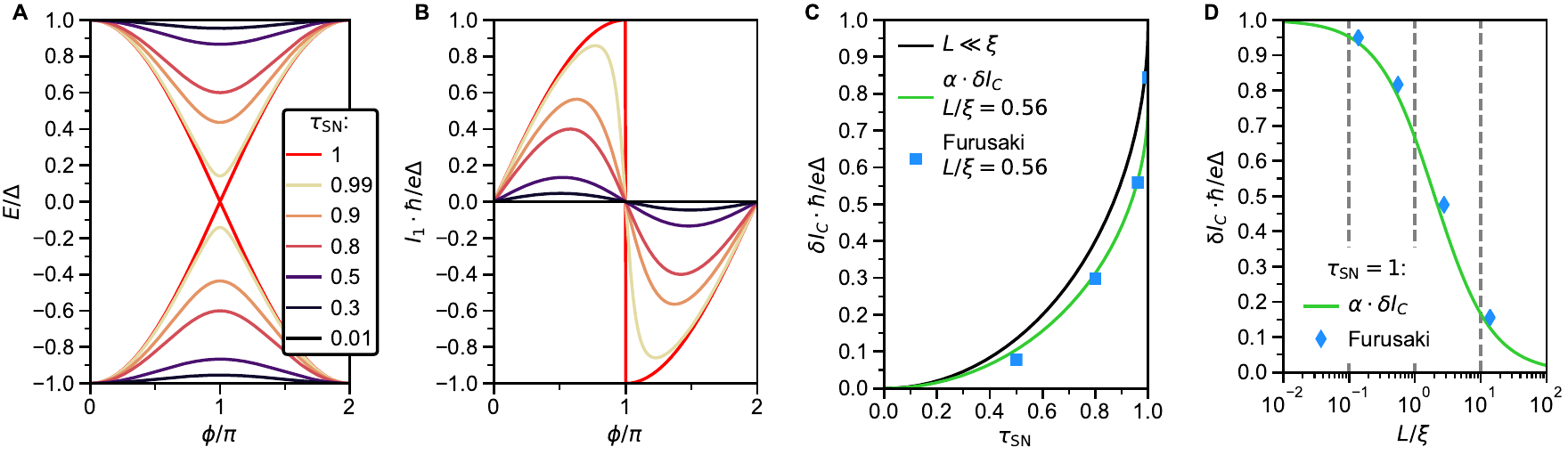}
\caption{\label{SM_JJ1ABS} short-limit SNS junction model: (\textbf{A}) Andreev bound state spectrum at different SN transparency levels, (\textbf{B}) current-phase relationship, also at different SN transparency levels, (\textbf{C}) critical current carried by a single ballistic mode, shown as a function of $\tau_\text{SN}$ ($\tau=\tau_\text{SN}^2$) in the short junction limit, and (\textbf{D}) junction length dependence in perfect transparency limit. Blue symbols in (\textbf{C}) and (\textbf{D}) are numerical results from \cite{Furusaki92}.}
\end{figure*}

The critical current of one mode is $\delta I_c = \text{max}(I_1(\phi))$. Fig.~\ref{SM_JJ1ABS}A-C shows the evolution of the ABS spectrum, the current-phase relationship and $\delta I_c$ with transparency.

\subsection{Critical current suppression with junction length}

For a long SNS junction \cite{Furusaki92, Bagwell92}, the Josephson current is carried by multiple ABS. With the length $L$ referring to the junction size along the current direction (Fig.~\ref{SM_sketch1}D), the number of ABS is approximately $L/\xi$. In the absence of scattering, the maximum supercurrent decreases as $1/L$. It was shown by Bagwell \cite{Bagwell92} that the crossover between the short ($L\ll\xi$) and long ($L\gg\xi$) limits can be interpolated as
\begin{equation}
\label{eqIcL1}
\delta I_c = \frac{e\Delta}{\hbar}\cdot\frac{1}{1+\frac{L}{2\xi}}.
\end{equation}

To treat the case of finite length and transparency, we adopt the approximation from \cite{Wan15} that the correction factor $\alpha$ for finite length is a multiplier for the current-phase relationship derived above for an SNS with finite transparency
\begin{equation}
\alpha=\frac{1}{1+\frac{L}{2\xi}},
\end{equation}

\begin{equation}
\label{eqIcL2}
\delta I_c = \text{max}(I_1(\phi))\cdot\alpha.
\end{equation}

As a cross-check, Fig.~\ref{SM_JJ1ABS}D illustrates that the equation~\ref{eqIcL1} is in agreement with a different calculation by Furusaki et al. \cite{Furusaki92}. We also verify that equation~\ref{eqIcL2} closely agrees with the calculation in \cite{Furusaki92} for the case of finite transparency at $L/\xi=0.56$ in Fig.~\ref{SM_JJ1ABS}C. We can see that a junction length smaller than but on the order of coherence length modestly suppresses the supercurrent.

\subsection{Saddle potential constriction model}
\label{sectionQPC}
The conductance plateau structure of a QPC is generally modeled by assuming a saddle potential profile \cite{Buttiker90}
\begin{equation}
V_\text{QPC}(x,y)=V_\text{QPC}(0,0)-\frac{m\omega_x^2 x^2}{2}+\frac{m\omega_y^2 y^2}{2},
\end{equation}
where $x$ and $y$ axes are parallel and orthogonal to the current flow, $V_\text{QPC}(0,0)$ is the potential at the center of the constriction, $\omega_x$ and $\omega_y$ describe the confining potential curvature. Transverse confinement discretizes the available states
\begin{equation}
E_n=V_\text{QPC}(0,0)+\left(n+\frac{1}{2}\right) \hbar\omega_y,
\end{equation}
with $n=$ 0, 1, 2, 3, ... The channels with transverse confinement energy below the Fermi energy are open, and above it are closed. The crossover between open and closed states is described by the transmission coefficient of an individual mode
\begin{equation}
T_n (E)=\frac{1}{1+\exp\left(\frac{-2\pi\left(E-E_n\right)}{\hbar\omega_x}\right)}.
\end{equation}

The split gate voltage tuning of the Fermi energy above the pinch-off voltage $V_P$ is generally described by a linear lever arm $C_\text{LA}$
\begin{equation}
T_n(V_\text{G12})=\frac{1}{1+\exp\left(V_P-C_\text{LA}V_\text{G12}-2\pi\left(n+\frac{1}{2}\right)\frac{\omega_y}{\omega_x}\right)}.
\end{equation}

Normal state conductance is then the sum across all available modes, each carrying a conductance quantum of $2e^2/h$ at full transmission
\begin{equation}
G_N(V_\text{G12}) = \frac{2e^2}{h}\cdot\sum_n T_n(V_\text{G12}),
\label{eqButtiker1}
\end{equation}
where the factor of 2 is from spin degeneracy.

In the extension of the saddle potential model to a ballistic SNS constriction, the critical current follows the same quantization pattern as $G_N$ \cite{Bauch05, Irie14}, but with a non-universal step height $\delta I_c$
\begin{equation}
I_c(V_\text{G12}) = \delta I_c \cdot \sum_n T_n(V_\text{G12}).
\label{eqButtiker2}
\end{equation}

For fitting a step structure in $I_c/\delta I_c$ or $G_N$ vs $V_\text{G12}$, the saddle potential model has three adjustable parameters: $V_P$ and $C_\text{LA}$ are used for position and rescaling on the $V_\text{G12}$ axis, and the confinement strength ratio $\omega_y/\omega_x$ for plateau sharpness. At high $\omega_y/\omega_x$ (long and narrow QPC), the discrete channel states are well separated in energy space, resulting in sharp, well defined plateaus \cite{Rossler11}.

\begin{figure}
\centering
\includegraphics[width=1.75in]{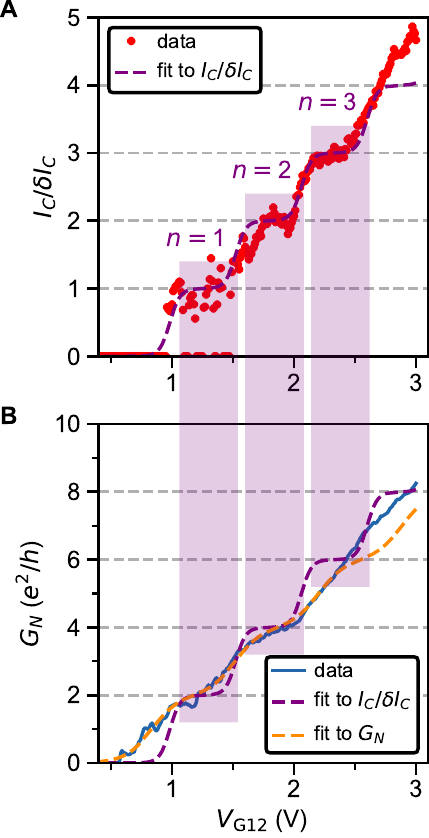}
\caption{\label{f2SM} Reproduction of Fig.~2C, D in the main text with a Buttiker model fit to $G_N$. (\textbf{A}) red circles is $I_c$ normalized to its step height for device 1A at $V_\text{GIL}=$ 3~V, $V_\text{BG} =$ 50~V, $T =$ 45~mK. (\textbf{B}) solid line is the normal state conductance at $V_\text{DC} = $ 100 $\mu$V, with a series resistance of 800 $\Omega$ subtracted from the raw data. The purple dashed lines in (\textbf{A}) is a fit of $I_c$ to eq.~\ref{eqButtiker2} with $\omega_y/\omega_x = 2.07$, and an arbitrary lever arm. The purple dashed line in (\textbf{B}) uses the same parameters to describe $G_N$ with eq.~\ref{eqButtiker1}. The orange dashed line in (\textbf{B}) is a fit of $G_N$ to eq.~\ref{eqButtiker1} in the $n=0-2.5$ region with $\omega_y/\omega_x = 0.99$.}
\end{figure}

Fig.~\ref{f2SM} reproduces Fig.3C and 3D in the main text, but with the saddle potential description of both $G_N$ and $I_c$ by equations~\ref{eqButtiker1} and \ref{eqButtiker2}. The purple dashed line in both figures is a fit to $I_c/\delta I_c$ using eq.~\ref{eqButtiker2} using values of $V_\text{G12} <$ 2.6~V, giving $\omega_y/\omega_x = 2.07$. The fit is very good for $I_c$ up to $n=3$ (Fig.~\ref{f2SM}A), but applying the same parameters to eq.~\ref{eqButtiker1} only approximately describes $G_N$ (Fig.~\ref{f2SM}B). The orange line in Fig.~\ref{f2SM}B is a fit to $G_N$  using values of $V_\text{G12} <$ 2.4~V, giving $\omega_y/\omega_x = 0.99$ and a good description of $G_N$ up to $n\approx2.5$.

We do not have a complete explanation of why the crossover between modes is sharper in $I_c$ then in $G_N$. The description above by a difference in confinement ratios suggests that the shape of the constriction is not the same  in the normal and superconducting states (that it is effectively narrower and/or longer in the latter). An alternate explanation could involve partial breakdown of the assumption that the  the constriction is adiabatically coupled to the leads \cite{Buttiker90,Mittag19}, and a different extent of this breakdown in the normal and superconducting states.

Another unusual aspect of Fig.~\ref{f2SM} is the occurrence of  $V_P$ at positive $V_\text{G12}$. Generically, pinch off is caused by depletion at negative split gate voltage \cite{Rossler11}. As documented further in section \ref{sectiondev2}, $V_P$ in our devices is a strong function of doping in the leads. It quickly shifts from positive at low electron density to negative at high density. A positive $V_P$ is consistent with a built-in depletion field at $V_\text{G12}=$ 0, for example from trapped charge at the gate metal/oxide interface \cite{Mikheev14}. A nominally positive $V_\text{G12}$ can thus still correspond to depletion around the split gates.

\subsection{OBTK model}
\label{sectionOBTK}
In this section, we briefly summarize the Octavio-Blonder-Tinkham-Klapwijk model \cite{Octavio83,Flensberg88} of an SNS constriction. It assumes a one-dimensional SNS weak link with two scattering barriers at each SN interface, with transparency $\tau_\text{SN}$. Its relationship to the equivalent SN barrier height is $\tau_\text{SN}=1/(1+Z_\text{SN}^2)$.

The current across the junction is calculated by integrating the distributions of right and left moving moving electrons ($f_\rightarrow$ and $f_\leftarrow$) in the energy space
\begin{equation}
\label{IVOBTK}
I=\frac{1}{e R_N} \int_{-\infty}^{+\infty} dE(f_\rightarrow(E)-f_\leftarrow(E)),
\end{equation}

\begin{equation}
f_\rightarrow(E)=A(E)f_\rightarrow(E-eV)+B(E)(1-f_\rightarrow(-E-eV))+T(E)f_0(E),
\end{equation}

\begin{equation}
f_\rightarrow(E)=f_\leftarrow(-E-eV),
\end{equation}
where $f_0$ is the standard Fermi function. At the SN interfaces, $A$ is the Andreev reflection probablity, $B$ is the ordinary reflection probablity, $T=1-A-B$ is the transmission probablity \cite{Blonder82}. For $E<\Delta$
\begin{equation}
A(E)=\frac{\Delta^2}{E^2+(\Delta^2-E^2)(1+2Z_\text{SN}^2)^2},
\end{equation}
\begin{equation}
B(E)=1-A(E).
\end{equation}

for $E>\Delta$:
\begin{equation}
A(E)=\frac{u^2_0 v^2_0}{\gamma^2},
\end{equation}
\begin{equation}
B(E)=\frac{(u^2_0-v^2_0)Z^2(1+Z_\text{SN}^2)}{\gamma^2}.
\end{equation}

The superconducting density of states $N_S$ enters the above equations as
\begin{equation}
N_S=\frac{1}{u_0^2-v_0^2},
\end{equation}

\begin{equation}
u_0^2=1-v_0^2=\frac{1}{2}\left(
1+\left(\frac{E^2-\Delta^2}{\Delta^2}\right)^{1/2}
\right),
\end{equation}

\begin{equation}
\gamma=u_0^2+Z_\text{SN}^2(u_0^2-v_0^2).
\end{equation}

\begin{figure}
\centering
\includegraphics[width=3.4in]{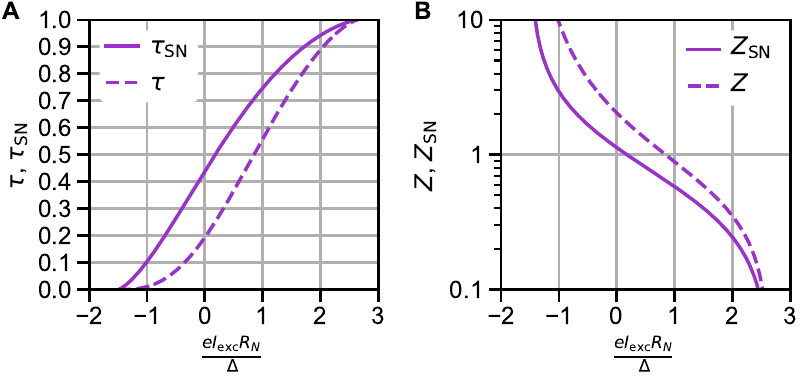}
\caption{\label{SM_OBTK}Mapping between excess current and transparency in the SNS model \cite{Octavio83,Flensberg88}: (\textbf{A}) SN and SNS transparency, (\textbf{B}) equivalent barrier heights.}
\end{figure}

The excess current $I_\text{exc}$ for a particular value of $\tau_\text{SN}$ can be found by calculating the $I(V)$ curve with eq.~\ref{IVOBTK}, and linearly extrapolating from $V\gg\Delta/e$ to $V=0$. The reverse mapping from dimensionless quantity $eI_\text{exc}R_N/\Delta$ to transparency is shown in Fig.~\ref{SM_OBTK}. This curve was used to estimate $\tau_\text{SN}$ from experimental $I_\text{DC}(V_\text{DC})$ curves (Fig.~3 in the main text and additional data shown in section~\ref{sectiondev123}).

\clearpage
\section{Additional devices and fabrication notes}

Sample 1: fabrication outlined in the methods section of the main text
\begin{itemize}
\item Device 1A: 40 nm gap between local gates. Data for a cooldown at $V_\text{GIL} = $ +3 V are discussed in the main text and throughout the supplementary material. 
\item Device 1B: 60 nm gap between local gates. Data for a cooldown at $V_\text{GIL} = $ +3 V are discussed in section~\ref{sectiondev123}. \end{itemize}

\begin{figure}[b]
\includegraphics[width=7in]{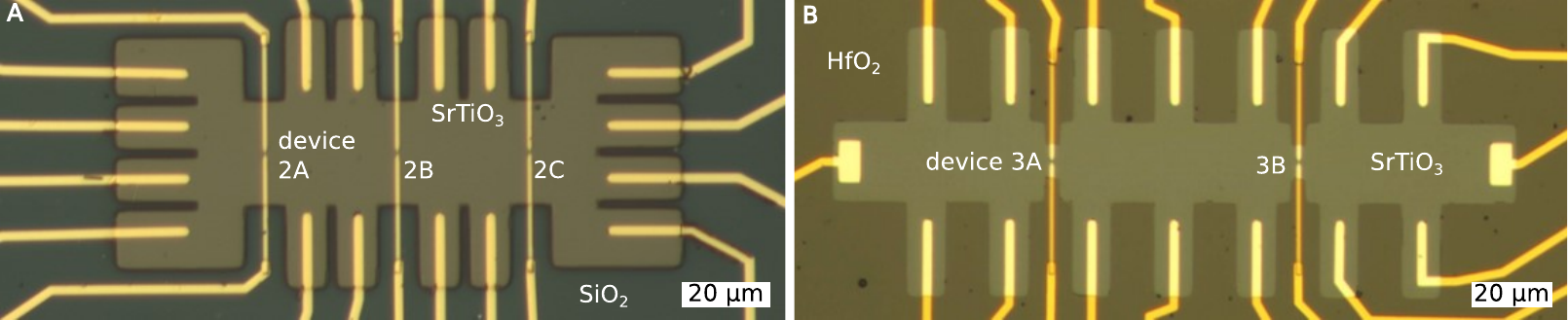}
\caption{\label{SM_dev23} Optical images of (\textbf{A}) sample 2 and (\textbf{B}) sample 3, taken at the end of the fabrication process. Optically visible gaps near the constriction are caused by the progressive narrowing of the gate tips.}
\end{figure}

\begin{figure*}
\centering
\includegraphics[width=3in]{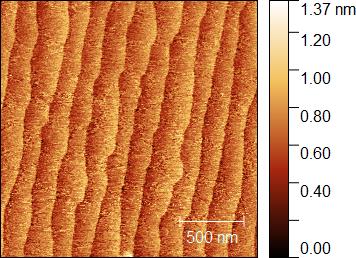}
\caption{\label{figS1} AFM image of atomic terrace steps on the surface of sample 1.}
\end{figure*}

Sample 2: same overall fabrication method as sample 1. Different design with three devices on the Hall bar with minor distinctions. The SiO$_2$ mesa insulator thickness was 100~nm and the final anneal before ionic liquid deposition was 1 minute at 180 \degree C. Optical image shown in Fig.~\ref{SM_dev23}A
\begin{itemize}
\item Device 2A: 40~nm gap between local gates. Data for cooldowns at $V_\text{GIL} = $ +3, +3.5 and +3.7 V are discussed in section~\ref{sectiondev2}.
\item Device 2B: 60~nm gap between local gates. Data for cooldowns at $V_\text{GIL} = $ +3, +3.5 and +3.7 V are discussed in section~\ref{sectiondev2}.
\item Device 2C: 100~nm gap between local gates, one of the gates was electrically shorted to the STO channel. 
\end{itemize}

Sample 3: same overall fabrication method as sample 1, but  sputtering of SiO$_2$ as mesa insulator in the last fabrication step is replaced with atomic layer deposition of thick HfO$_2$. The lift-off procedure was similar to the local gate patterning in step 1, but with 200 cycles of atomic layer deposition. The final anneal was 20 minutes at 115 \degree C. Optical image shown in Fig.~\ref{SM_dev23}B. We found this alternative approach to depositing mesa insulators to be viable but detrimental to ohmic contacts, which suffered from poor yield and were only functional at relatively high carrier densities.
\begin{itemize}
\item Device 3A: 40~nm gap between local gates. Data for a cooldown at $V_\text{GIL} = $ +3~V are discussed  in section~\ref{sectiondev123}.
\item Device 3B: 60~nm gap between local gates. Data for a cooldown at $V_\text{GIL} = $ +3~V are discussed in section~\ref{sectiondev123}.
\end{itemize}

Fig.~\ref{figS1} shows an atomic force microscope (AFM) image of SrTiO$_3$ surface. Atomic terrace steps due to surface miscut are clearly visible. The image was taken on sample 1 after surface preparation with a deionized water soak, followed by an anneal at 1000 \degree C in an Ar/O$_2$ atmosphere. It was taken prior to fabrication of devices 1A and 1B on this sample.

\begin{figure}
\includegraphics[width=7in]{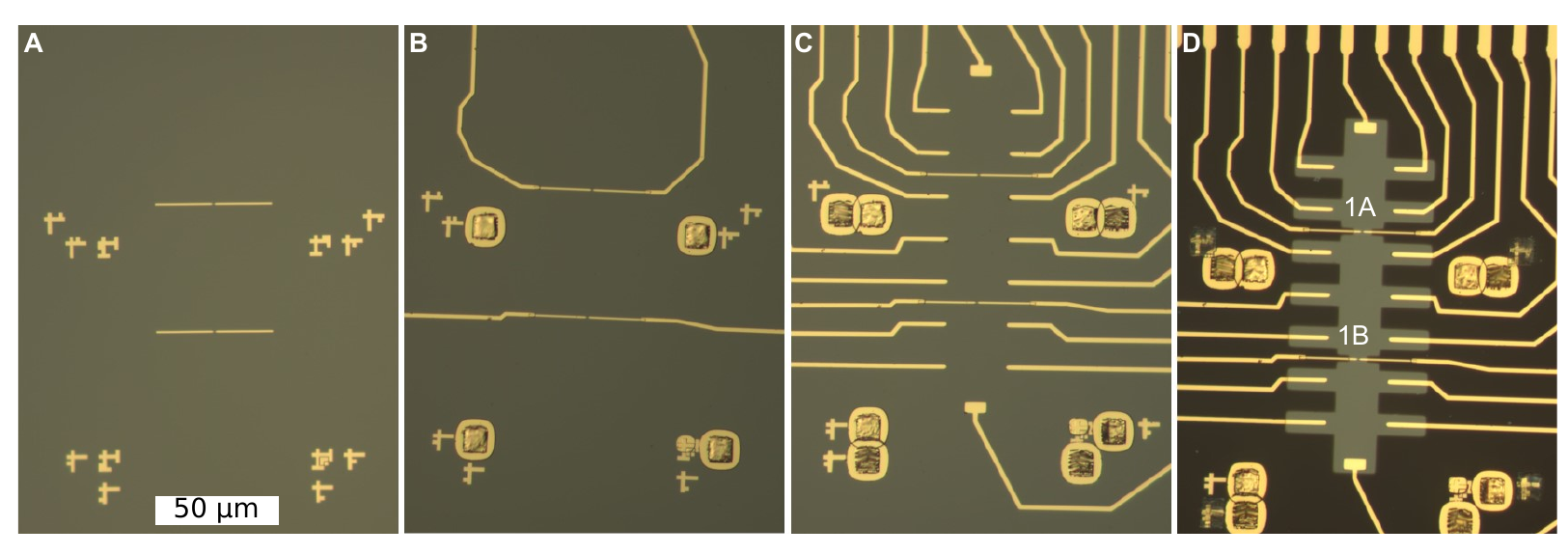}
\caption{\label{SM_dev1A} Optical images of devices 1A and 1B taken at intermediate stages of the fabrication process: (\textbf{A}) gate lift-off, (\textbf{B}) gate contact lift-off, (\textbf{C}) ohmic contact lift-off, (\textbf{D}) mesa insulation lift-off.}
\end{figure}

\begin{figure}
\includegraphics[width=3in]{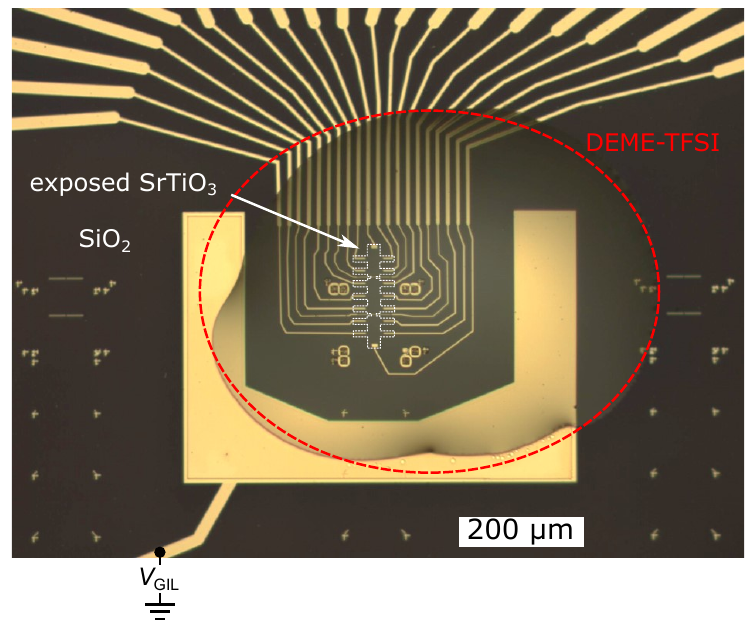}
\caption{ Optical image of sample 1, taken after the measurements in a dilution fridge, with the ionic liquid covering devices 1A, 1B, and the large coplanar gate.}
\label{SM_dev1A2}
\end{figure}

Intermediate stages of the fabrication process are illustrated  in Fig.~\ref{SM_dev1A}. The presented optical images are centered around the Hall bar-style mesa that is defined in the final step by depositing SiO$_2$ insulation. A larger view of the device is shown in Fig.~\ref{SM_dev1A2}, which includes the ionic liquid and the large coplanar gate used to drive an insulator-to-metal transition on the exposed SrTiO$_3$ surface.

\clearpage

\section{Tuning superconductivity in the leads with ionic liquid gating}

\begin{figure}
\includegraphics[width=7in]{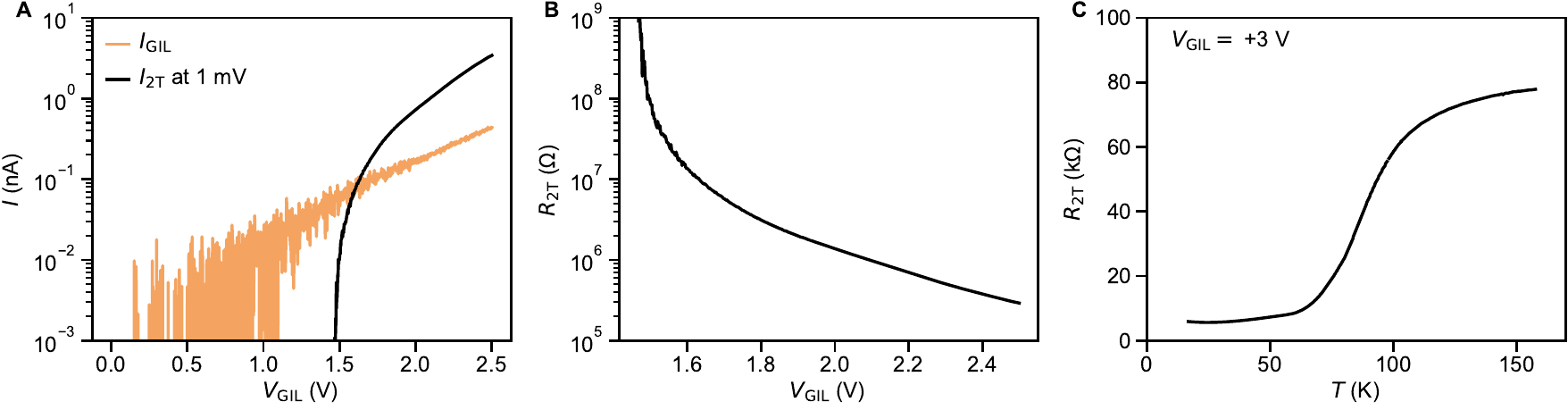}
\caption{\label{figS_VgIL2} (\textbf{A}) Initial sweep of the large coplanar gate $V_\text{GIL}$ measured for device 3 at room temperature. Channel current $I_\text{2T}$ and coplanar gate leakage $I_\text{GIL}$ are shown. (\textbf{B}) Same sweep presented as channel  resistance $R_\text{2T}$. (\textbf{C}) $R_\text{2T}$ measured during cooldown at fixed $V_\text{GIL} =$ 3~V.}
\end{figure}

Fig.~\ref{figS_VgIL2} illustrates the initial steps of the device measurement. The initial ramping of the coplanar gate ($V_\text{GIL}$) is performed at room temperature and is monitored by a two terminal-like measurement. A DC voltage of 1~mV is sourced to one chosen ohmic contact and  all remaining contacts on the Hall bar are grounded. The resulting current $I_\text{2T}$ typically becomes measurable near $V_\text{GIL} =$ 1 - 2 V and quickly increases by several orders of magnitude. The corresponding resistance $R_\text{2T}$ typically saturates around 100~k$\Omega$. It includes contributions from diffusive scattering in the SrTiO$_3$ channel, contact and line resistances. The dominant contributions at room temperature are channel and contact resistances.

If the carrier density induced in the channel is sufficiently large to make it metallic ($N>5\cdot 10^{12}$~cm$^{-2}$), the measured resistance quickly decreases upon cooling. $R_\text{2T}$ becomes dominated by contact and line resistances at low temperatures. During low temperature measurements, $V_\text{GIL}$ is kept at a fixed value chosen at the start of the cooldown. Below 220 K, the ionic liquid is frozen and does not respond to adjustments of $V_\text{GIL}$. To re-adjust the carrier density in the Hall bar by changing $V_\text{GIL}$, the device needs to be thermally cycled above that point, as was done for device 2.

\begin{figure*}
\centering
\includegraphics[width=7in]{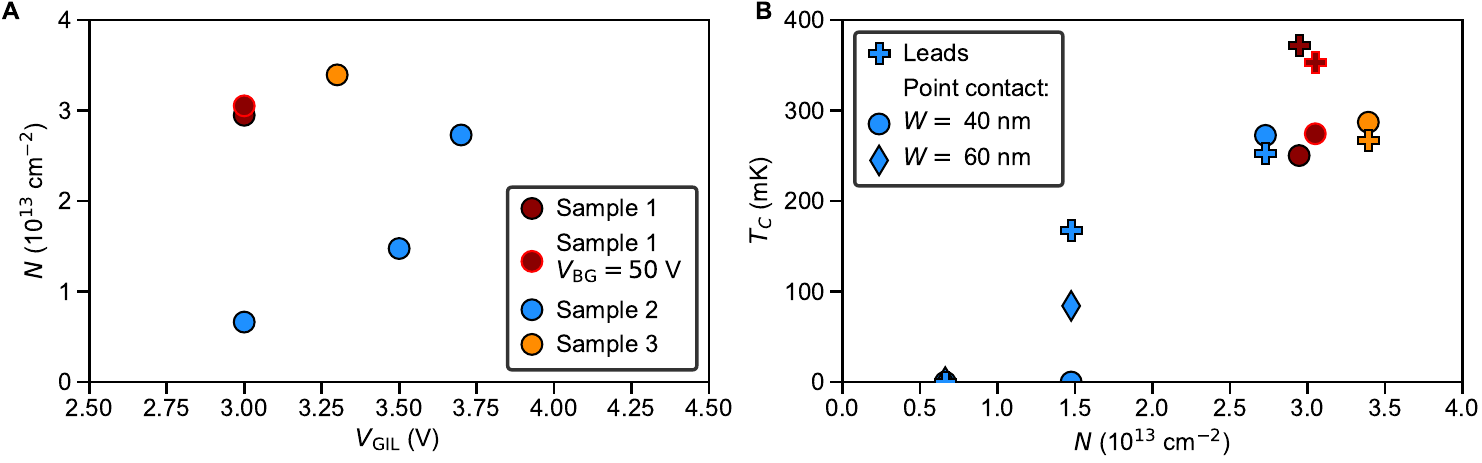}
\caption{\label{figSM_VgIL} Summary of different devices and cooldowns. (\textbf{A}) low-field Hall density measured in the leads at base temperature. For sample 1, the effect of applying a back voltage $V_\text{BG} =$ +50 V is indicated. $V_\text{BG} =$ 0 V for other points. (\textbf{B}) Superconducting transition as indicated by the midpoint of resistance drop. Symbol markers indicate measurement in the leads vs across the point contact. Colors correspond to different samples.}
\end{figure*}

Fig.~\ref{figSM_VgIL} summarizes the various cooldowns performed on samples 1-3. While the trend of Hall density with $V_\text{GIL}$ shows  significant scatter between samples, it is monotonic for successive cooldowns on the same sample (sample 2). The superconducting transition points shown in Fig.~\ref{figSM_VgIL}B are defined as the midpoint of the resistance drop measured in the leads or across the gated constriction. For constrictions, transition temperatures were extracted in ‘‘open’’ state at large positive local gate voltage. The results are consistent with a dome-shaped superconducting phase, with an onset of $T_c$ at carrier densities above $1\cdot10^{13}$~cm$^{-2}$ and a peak near $3\cdot10^{13}$~cm$^{-2}$.

This work focuses on the underdoped and near optimal regimes, where the carrier density is low enough that local constrictions are electrostatically tunable by HfO$_2$ dielectric gates.

A likely source of uncertainty in extracted carrier densities is multiple band occupancy resulting in non-linear Hall effect. Strong non-linearity of transverse resistance with magnetic field of is commonly observed at high carrier density in ionic liquid-gated SrTiO$_3$ and LAO/STO \cite{Gallagher15, Joshua12}. However, only weak deviation from linearity was observed up to 14~T for device 1 and up to 3~T for device 3.

\clearpage
\section{Additional characterization of the local and back gates}

\begin{figure*}[b]
\centering
\includegraphics[width=7in]{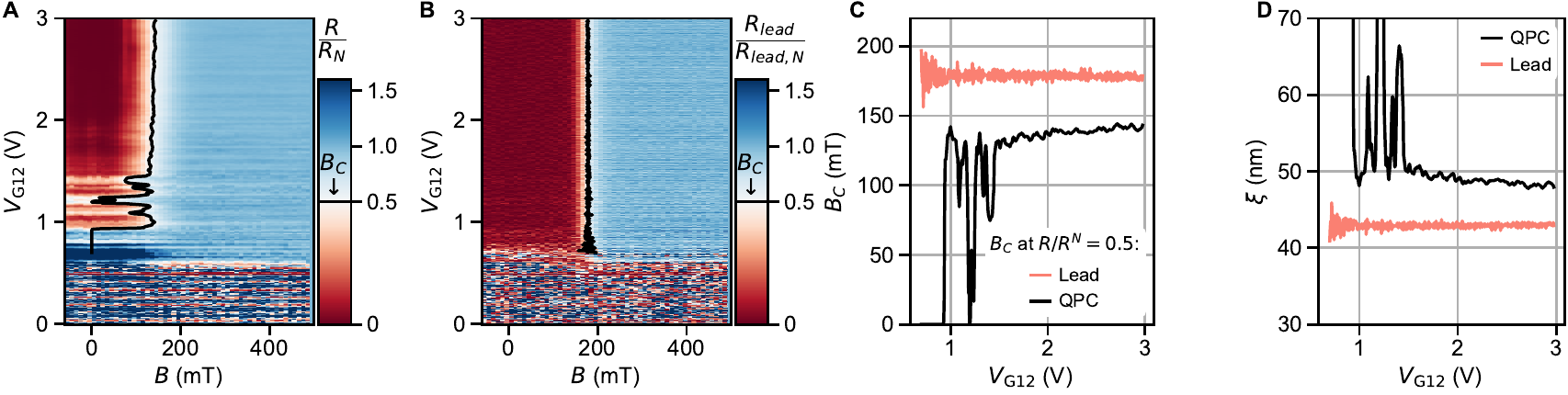}
\caption{\label{SM_Bc} Normalized resistance of (\textbf{A}) the constriction and (\textbf{B}) the lead as a function of magnetic field and local gate voltage in device 1A at $V_\text{BG} =$ 50~V, $V_\text{GIL} =$ 3~V, $T =$ 44~mK. Normal state resistance $R_N$ is taken $B =$ 500 mT. Black lines indicate the critical field $B_c$, defined as the midpoint of the resistance drop. Critical fields extracted in (\textbf{A}) and (\textbf{B}) are re-plotted in (\textbf{C}). (\textbf{D}) Superconducting coherence length estimate extracted from $B_c$ using equation \ref{eqBc}.}
\end{figure*}

Fig.~\ref{SM_Bc} illustrates the extraction of the Ginzburg-Landau coherence length estimate $\xi$ from the critical field $B_c$ of the constriction and the leads with
\begin{equation}
\xi^{2}=\frac{\Phi_0}{2\pi B_\text{C}},
\label{eqBc}
\end{equation}
with $\Phi_0$ being the flux quantum. Using the criterion $B_c = B(R/R_N = 0.5)$, with $R_N$ being the normal state resistance at high $B$, $\xi =$ 43~nm in the leads. In the constriction,  $\xi$ decreases from 50 to 48~nm with increasing $V_\text{G12}$. If one chooses a lower criterion $B_c = B(R/R_N = 0.25)$, the slightly modified estimates are: 45~nm in the leads and 53-60~nm in the constriction. As a function of $V_\text{G12}$, the $B_c$ measurement in the leads remains unchanged until the constriction is closed near $V_\text{G12} =$ 0.8~V and there is no sourcing current. In the constriction measurement, the supercurrent becomes intermittent below $V_\text{G12} =$ 1.5~V due to the resonant nature of barrier transmission in this regime. The spikes in $B_c$ and $\xi$ extracted at low $V_\text{G12}$ are thus not necessarily reflective of any actual change in the superconducting order near the constriction.

\begin{figure*}
\centering
\includegraphics[width=7in]{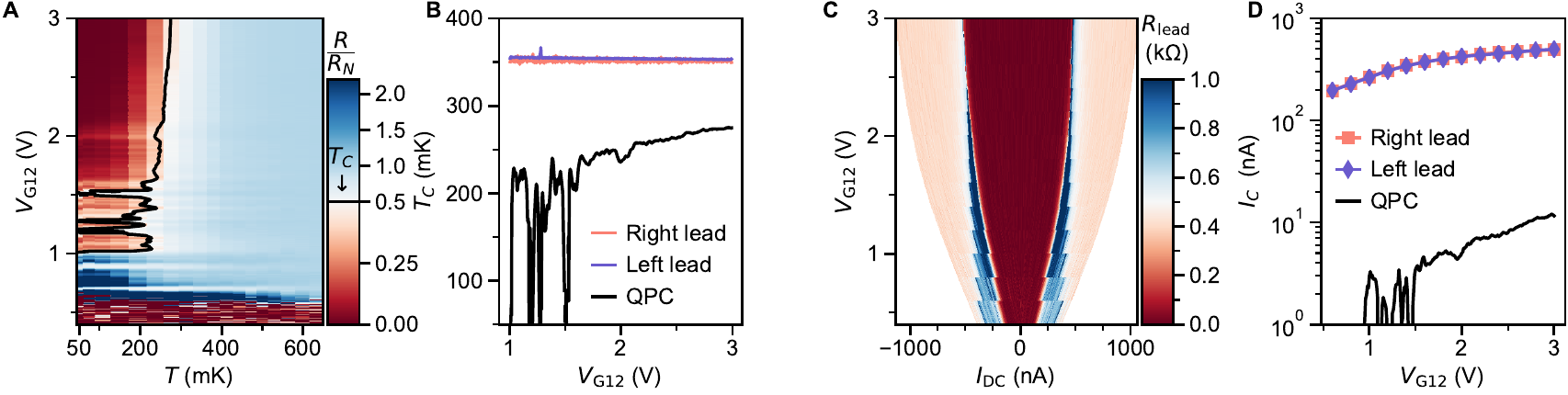}
\caption{\label{SM_Vg12} (\textbf{A}) Normalized resistance of the constriction  as a function of temperature and local gate voltage in device 1A at $V_\text{BG} =$ 50~V, $V_\text{GIL} =$ 3~V. (\textbf{B}) $T_c$ extracted in (\textbf{A}) and from concurrent measurement of lead resistance. (\textbf{C}) Lead resistance as a function of DC current and local gate voltage at $V_\text{BG} =$ 50~V, $V_\text{GIL} =$ 3~V, $T =$ 44~mK. (\textbf{D}) Comparison between the critical current in the leads and in the constriction.}
\end{figure*}

Fig.~\ref{SM_Vg12}A shows the extraction of $T_c$ from the constriction resistance measurement, where similar intermittency is seen at low $V_\text{G12}$. While the $T_c$ of the constriction is tuned between 200 and 275~mK by the local gate, $T_c$ in the leads remains constant at 350~mK.

Interestingly, the only measurement in which the leads are sensitive to $V_\text{G12}$ is their critical current. Fig.~\ref{SM_Vg12}C shows the lead resistance up to high DC current at base temperature, where a decreased $I_c$ is clearly seen at low $V_\text{G12}$. This suggests that the local gate can also have a very long range effect on the leads, with the closest voltage probe being 5 microns away. Such an effect is plausible given the highly non-linear dielectric constant of SrTiO$_3$, and a presumably non-uniform carrier density profile in the Hall bar, with increased depletion at the edges \cite{Persky20}. Nevertheless, Fig.~\ref{SM_Vg12}D illustrates that $I_c$ in the leads  remains two orders of magnitude above $I_c$ in the constriction.

\begin{figure*}
\centering
\includegraphics[width=5.3in]{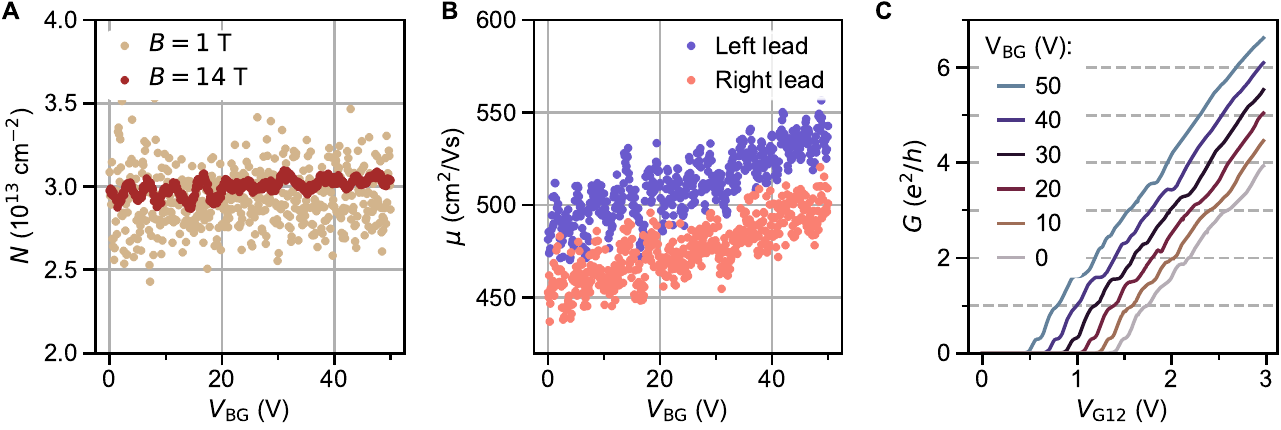}
\caption{\label{SM_Vbg}(\textbf{A}) Carrier density dependence on back gate voltage, extracted from Hall effect measured in the low and high field limits for device 1A at $V_\text{G12} =$ +3~V, $V_\text{GIL} =$ 3~V, $T =$ 45~mK. (\textbf{B}) Hall mobility in the leads, calculated from the high-field limit value of $N$. (\textbf{C}) Constriction conductance traces with local gate voltage, at fixed $V_\text{BG} =$ 0-50~V, taken in the normal state at $T = $ 866~mK. Data is presented without offset shifting and series resistance correction.}
\end{figure*}

\begin{figure*}
\centering
\includegraphics[width=7in]{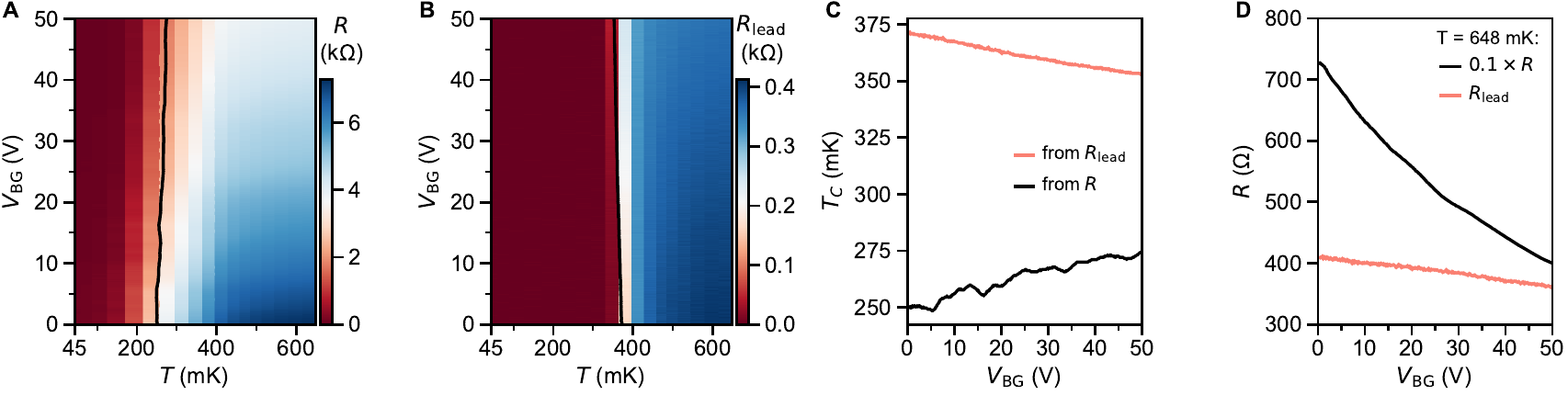}
\caption{\label{SM_TcVbg} Back gate tuning of $T_c$ in device 1A at $V_\text{G12} =$ 3~V, $V_\text{GIL} =$ 3~V . (\textbf{A}) Constriction and (\textbf{B}) lead resistance as a function of temperature and back gate voltage. Solid black lines in (\textbf{A},\textbf{B}) indicate $T_c$. $T_c$ values extracted in (\textbf{A}) and (\textbf{B}) are re-plotted in (\textbf{C}). (\textbf{D}) Constriction and lead resistance measured above $T_c$.}
\end{figure*}

\begin{figure*}
\centering
\includegraphics[width=7in]{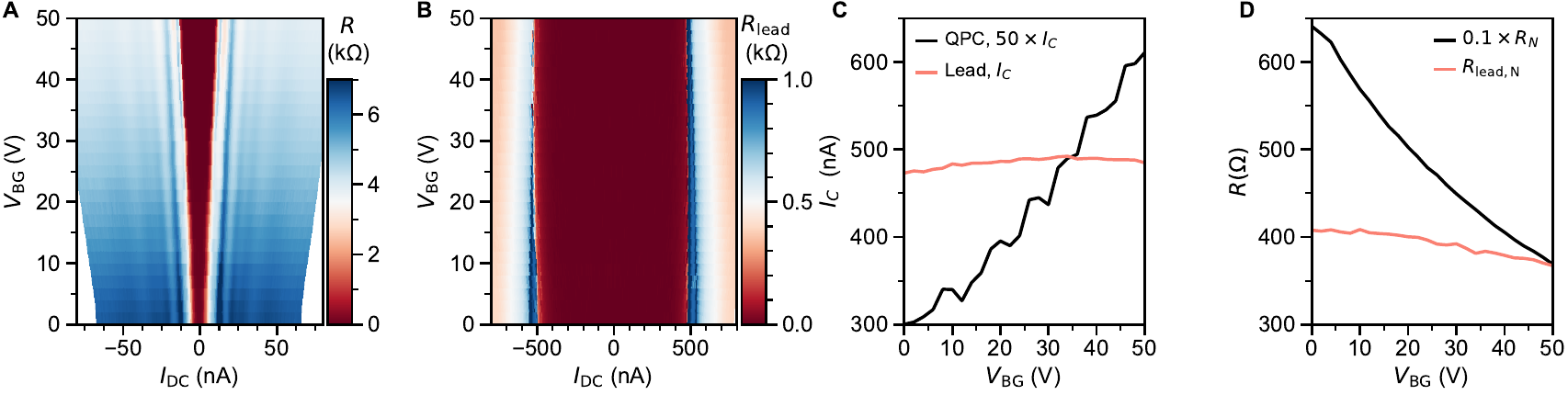}
\caption{\label{SM_IcVbg}Back gate tuning of critical current $I_c$ in device 1A at $V_\text{G12} =$ 3~V, $V_\text{GIL} =$ 3~V, $T =$ 44~mK. (\textbf{A}) Constriction and (\textbf{B}) lead resistance as a function of DC source current and back gate voltage. Black lines indicate $T_c$. (\textbf{C}) Critical current values for the resistance drop in (\textbf{A}) and (\textbf{B}). (\textbf{D}) Normal state constriction and lead resistance taken at high DC current.}
\end{figure*}

Back gate voltage $V_\text{BG}$ is an additional tuning knob for our device. $V_\text{BG}$ is applied between the device and the bottom surface of a 0.5 mm thick SrTiO$_3$ crystal, connected to the bottom of a chip carrier with silver paint. Such gates can have an appreciable capacitance due to the quantum paraelectric nature of SrTiO$_3$, resulting in a dielectric constant of order 10$^4$ in the low temperature limit \cite{Honig13, Chen16, Persky20}.

Unlike $V_\text{G12}$, $V_\text{BG}$ affects both the constriction and the leads. Similarly to many previous experiments on LAO/STO 2DESs \cite{Bell09, Chen16}, the main effect of $V_\text{BG}$ is to change the carrier mobility $\mu = 1/(eNR_\text{lead})$, rather than to change the density $N$. The back gate effect in  Fig.~\ref{SM_Vbg} is 3\% on $N$ and 10\% on $\mu$. This is consistent with the understanding that positive $V_\text{BG}$  pulls the 2DES away from disorder scattering at the surface, increasing the mobility \cite{Chen16}. This work has thus mainly focused on the $V_\text{BG}$ = 50~V state, which offers the highest 2DES mobility. In our devices, the capability to deplete using $V_\text{BG}$ is limited due to rapid damage suffered by ohmic contacts upon applying negative $V_\text{BG}$. Such damage is largely reversible upon thermally cycling the device to near room temperatures, suggesting a charge trapping mechanism similar to the one documented in \cite{Seri13,Biscaras14}. Similar contact damage can occur at negative $V_\text{G12}$, particularly at low $V_\text{BG}$. This is likely a consequence of capacitive cross-coupling between the gates.

Fig.~\ref{SM_TcVbg} shows the modulation  of the superconducting transition by $V_\text{BG}$. The lead $T_c$ is decreased from 370~mK to 350~mK by increasing $V_\text{BG}$. This is consistent with a near-optimal, slightly overdoped position on the the superconducting dome.

The back gate effect on the constriction resistance is proportionally much larger than for the leads. This is shown in one dimensional sweeps of $V_\text{BG}$ in the normal state at above $T_c$ in Fig.~\ref{SM_TcVbg}D and at high DC bias in Fig.~\ref{SM_IcVbg}D. A comparison of $V_\text{G12}$ sweeps at different $V_\text{BG}$ (Fig.~\ref{SM_Vbg}D) show that the most obvious effect is a horizontal shift in $V_\text{G12}$, suggesting a cross-coupling effect. Beyond the horizontal shift, there is a vertical shift of most non-monotonic features in $G$, such as the short plateaus  near $2e^2/h$ and $4e^2/h$ that are most clearly visible at $V_\text{BG} =$ 50~V. The downward trend of such features suggests an increased series resistance at low $V_\text{BG}$, consistent with increased $R_\text{lead}$. For further discussion of plateau features in the normal state and series resistance correction, see section \ref{sectiondev2}.

Fig.~\ref{SM_IcVbg} shows the superconducting critical current $I_c$ of an open constriction ($V_\text{G12} =$ 3~V). Back gate voltage has a strong effect on normal state resistance. While the simultaneous lack of a strong trend in $I_c$ of the lead is in apparent conflict with the trend in $T_c$, similar trends (optimal $I_c$ at higher than optimal doping $T_c$) have been observed in LAO/STO Hall bars \cite{Prawiroatmodjo16,Hurand19}.

\clearpage
\section{Normal state conductance of the constriction, evolution from normal to superconducting leads}
\label{sectiondev2}

\subsection{Sample 2}
\begin{figure*}[b]
\centering
\includegraphics[width=7in]{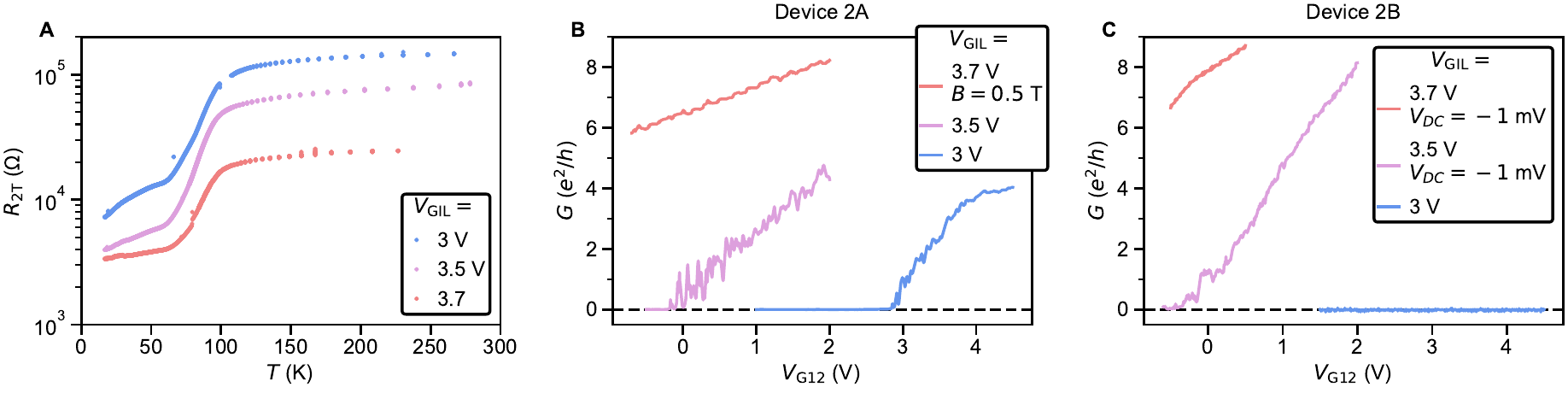}
\caption{\label{SM_dev2_Vg12} Sample 2, normal state conductance of devices (\textbf{A}) 2A and (\textbf{B}) 2B at base temperature for three cooldowns at different $V_\text{GIL}$. Raw data without $R_S$ subtraction is shown.}
\end{figure*}

Sample 2 was studied across three cooldowns at $V_\text{GIL} =$ 3, 3.5, and 3.7~V. Larger $V_\text{GIL}$ increases the carrier density in the Hall bar channel (Fig.~\ref{figSM_VgIL}A), and drives it towards more robust metallicity  (Fig.~\ref{SM_dev2_Vg12}A) and superconductivity  (Fig.~\ref{figSM_VgIL}B). Fig.~\ref{SM_dev2_Vg12}B and \ref{SM_dev2_Vg12}C illustrate the evolution of constriction conductance behavior with $V_\text{GIL}$ and split gate voltage $V_\text{G12}$, at base temperature (28-35~mK). Increased metallicity in the channel at higher $V_\text{GIL}$ translates into a rapid shift of the constriction pinch off point to lower $V_\text{G12}$.

\begin{figure*}
\centering
\includegraphics[width=5.3in]{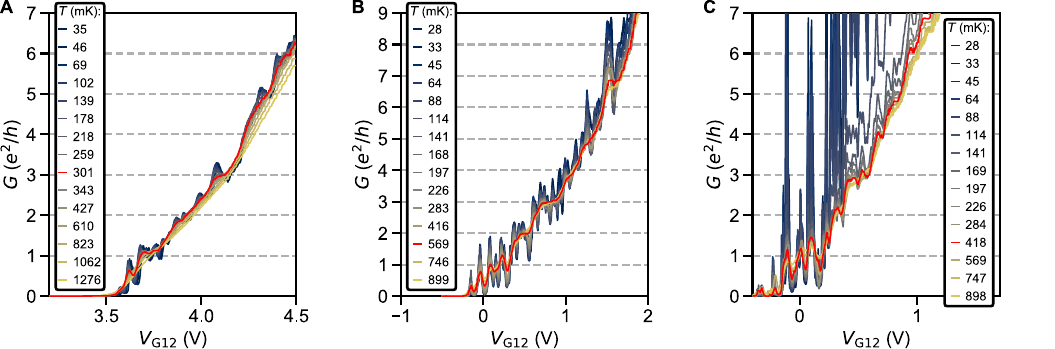}
\caption{\label{SM_dev2_Vg12T} Sample 2, temperature dependence of constriction conductance at zero DC bias. The red color emphasizes a trace at an intermediate $T$, which reduces the strength of low-$T$ fluctuations but remains below the onset of strong thermal smearing. (\textbf{A}) device 2A, $V_\text{GIL} = $ 3~V, (\textbf{B}) device 2A, $V_\text{GIL} = $ 3.5~V, (\textbf{C}) device 2B, $V_\text{GIL} = $ 3.5~V. A series resistance $R_S$ = 3.5~k$\Omega$ was subtracted in (\textbf{A}, \textbf{B}) and 1.4~k$\Omega$ in (\textbf{C}).}
\end{figure*}
\begin{figure*}
\centering
\includegraphics[width=3.6in]{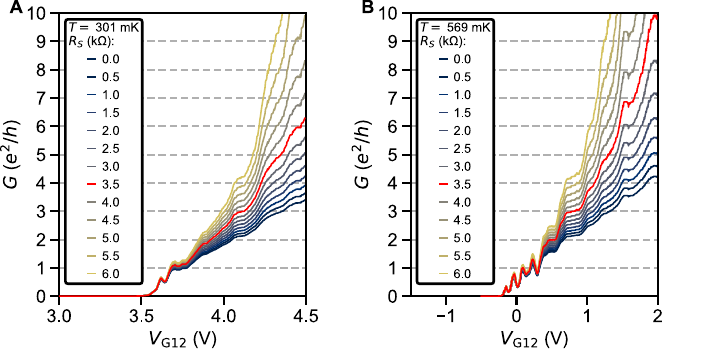}
\caption{\label{SM_dev2_Rc} Illustration of series resistance subtraction from conductance-split gate voltage traces for device 2A at (\textbf{A}) $V_\text{GIL}=$ 3~V, 301~mK, (\textbf{B})  $V_\text{GIL}=$ 3.5~V, 569~mK. $R_S = $ 0 corresponds to the raw data. The red traces correspond to the ultimately chosen value of $R_S = $ 3.5~k$\Omega$.}\end{figure*}

At $V_\text{GIL} =$ 3~V, the 60~nm wide constriction (device 2B) remains closed up to $V_\text{G12} = $ 4.5~V. The 40~nm wide constriction (device 2A) becomes open near $V_\text{G12} = $ 3~V. The occurrence of pinch-off at positive $V_\text{G12}$ despite metallic conductivity in the leads can be understood in terms of a built-in depletion field around constriction edges. At $V_\text{GIL} =$ 3.5 and 3.7~V, the Hall bar channel becomes superconducting at base temperature. In both cases, the constriction in device 2B also becomes superconducting in its open state at large $V_\text{G12}$. The constriction in the device 2A does not show a clear supercurrent at $V_\text{GIL} =$ 3.5~V, but does become superconducting at 3.7~V. In presence of a supercurrent, the traces shown in Fig.~\ref{SM_dev2_Vg12} are taken in the normal state in magnetic field or at high DC bias.

Fig.~\ref{SM_dev2_Vg12T} shows constriction conductance traces with $V_\text{G12}$ at different temperatures. Common trends for both devices 2A and 2B, $V_\text{GIL} = $ 3 and 3.5 V are: plateau signatures near both even and odd multiples of $\delta G = e^2/h$, fluctuations of $G$, repeatable between $V_\text{G12}$ sweeps, with amplitude decreasing with temperature. Quantization is best seen at an intermediate $T$, where fluctuations are reduced, but the thermal smearing is not yet fully onset. This highlighted by selected red line traces in Fig.~\ref{SM_dev2_Vg12T}.

\begin{figure*}
\centering
\includegraphics[width=7in]{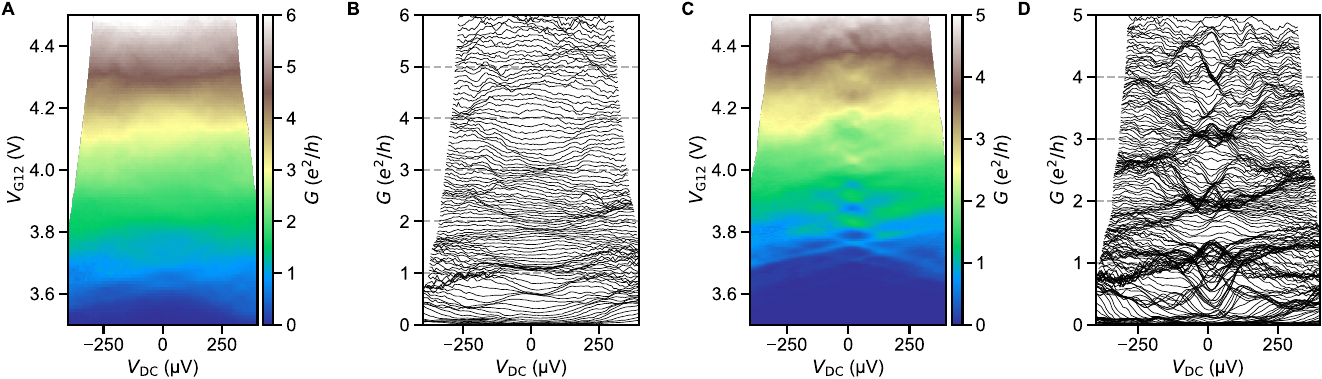}
\caption{\label{SM_dev2_Vg12Vdc_1} Constriction conductance map and corresponding line traces with $V_\text{G12}$ and DC bias. Device 2A, $V_\text{GIL} = $ 3~V at (\textbf{A}, \textbf{B}) 302~mK and (\textbf{C}, \textbf{D}) 35~mK. A series resistance $R_S$ = 3.5~k$\Omega$ was subtracted in (\textbf{A}-\textbf{D}).}
\end{figure*}

\begin{figure*}
\centering
\includegraphics[width=7in]{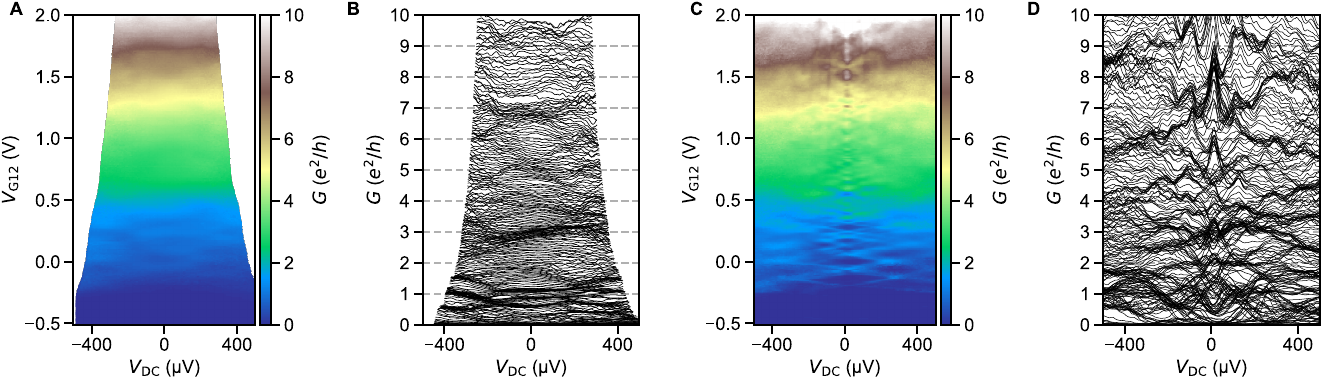}
\caption{\label{SM_dev2_Vg12Vdc_2} Constriction conductance map and corresponding line traces with $V_\text{G12}$ and DC bias. Device 2A, $V_\text{GIL} = $ 3.5~V at (\textbf{A}, \textbf{B}) 751~mK and (\textbf{C}, \textbf{D}) 28~mK. A series resistance $R_S$ = 3.5~k$\Omega$ was subtracted in (\textbf{A}-\textbf{D}).}
\end{figure*}

\begin{figure*}
\centering
\includegraphics[width=7in]{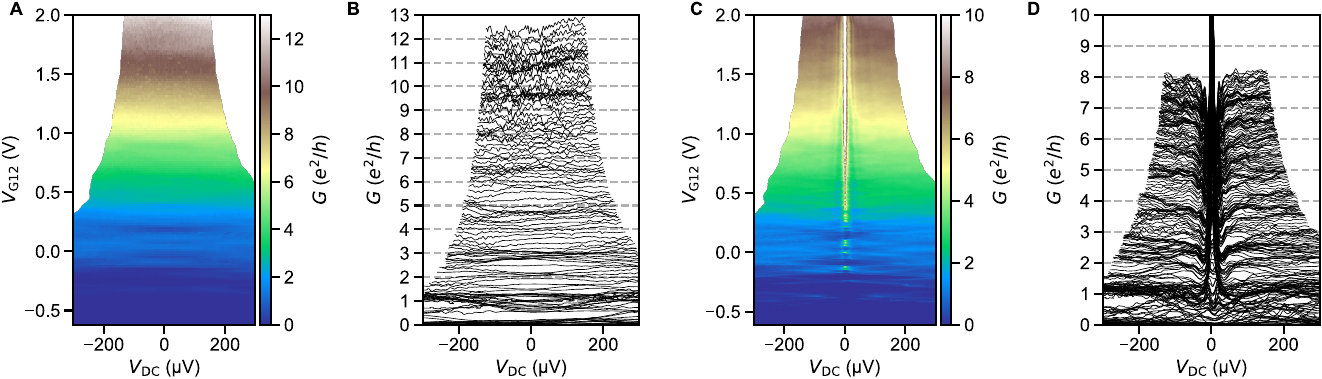}
\caption{\label{SM_dev2_Vg12Vdc_3} Constriction conductance map and corresponding line traces with $V_\text{G12}$ and DC bias. Device 2B, $V_\text{GIL} = $ 3.5~V at (\textbf{A},\textbf{B}) 302~mK and (\textbf{C}, \textbf{D}) 28~mK. A series resistance $R_S$ = 1.4~k$\Omega$ was subtracted in (\textbf{A}, \textbf{B}) and $R_S$ = 0 in (\textbf{C}, \textbf{D}).}
\end{figure*}

Additional evidence for conductance quantization is seen in DC bias spectroscopy in Fig.~\ref{SM_dev2_Vg12Vdc_1}, \ref{SM_dev2_Vg12Vdc_2} and \ref{SM_dev2_Vg12Vdc_3}. Line traces with $V_\text{DC}$ at different $V_\text{G12}$ tend to crowd around multiples of $e^2/h$ at $V_\text{DC} = 0$. This is again most clearly seen at a $T$ slightly elevated from base, and it is obscured at low $T$ by fluctuations. At finite $V_\text{DC}$, traces crowd at half values between multiples of $e^2/h$ (most clearly seen in Fig.~\ref{SM_dev2_Vg12Vdc_1}A). This is consistent with the classic picture of DC bias adding an extra available ballistic mode for carriers moving in one direction only \cite{Kouwenhoven89}.

Regarding conductance quantization in steps of $\delta G = e^2/h$ (as opposed to $\delta G = 2e^2/h$), it is difficult to unambiguously disentangle the series contact resistance contribution. In this paper, we adopt the standard simple approach of subtracting a constant value $R_S$ from the measured constriction resistance $R$ (Fig.~\ref{SM_dev2_Rc}). Identification of the lower conductance plateaus, particularly at the first two multiples of $e^2/h$, is only weakly affected by the arbitrary choice of $R_S$. However, the absolute value of higher conductance plateaus is quite sensitive to small adjustments in $R_S$, and can thus easily be misidentified. Additionally, it is unclear whether in our device geometry $R_S$ is truly independent of $V_\text{G12}$. One reasonable scenario is a reduction of $R_S$ at high $V_\text{G12}$ by carrier accumulation in the STO regions neighboring the constriction. One hint at challenges with constant $R_S$ subtraction is the absence of plateau signatures at $4e^2/h$ in Fig.~\ref{SM_dev2_Vg12T}A and $6e^2/h$ in Fig.~\ref{SM_dev2_Vg12T}B. Nevertheless, the data in Fig.~\ref{SM_dev2_Vg12T} is clearly more compatible with a pattern of multiples in $e^2/h$ rather than $2e^2/h$, particularly given fairly robust features at $G = e^2/h$ and $3e^2/h$.

Conductance quantization with $\delta G = e^2/h$ is expected in any QPC when the spin degenaracy is lifted by a magnetic field $B$ \cite{Buttiker90}. However, even in the absence of $B$, such half-quantization has been reported in many studies of gated constrictions and quantum wires based on InAs \cite{Debray09,Matsuo17,Mittag19}, GaAs \cite{Crook06,Hew08,Scheller14}, and carbon nanotubes \cite{Biercuk05}. The precise origins of this effect have arguably not been elucidated \cite{Matsuo17,Mittag19}. One ingredient suspected to be important is the presence of either intrinsic or field-induced spin-orbit interaction, resulting in spontaneous spin polarization \cite{Wan09} or filtering transmission for opposite spins \cite{Kohda12}. Another possible essential ingredient is electron-electron interaction in 1D confinement, creating a spin incoherent or helically ordered Luttinger liquid  \cite{Matveev04,Fiete07}. Another proposed mechanism is disruption of the adiabaticity of the constriction-lead coupling by the disorder potential \cite{Mittag19}. All of these mechanisms are potentially relevant for the case of a narrow constriction in STO. Quantization with $\delta G = e^2/h$ has also been reported for STO in accidental QPC's in shorted line junctions \cite{Gallagher14} and LAO/STO wires \cite{Ron14}.

\clearpage
\subsection{Sample 1}

\begin{figure*}[b]
\centering
\includegraphics[width=5.3in]{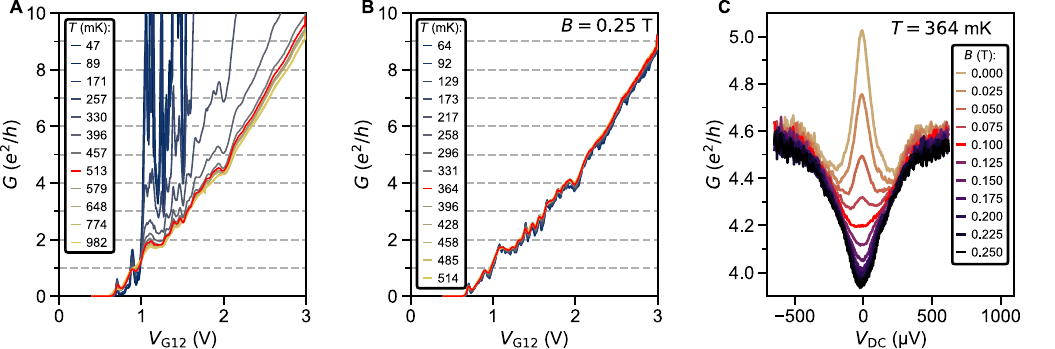}
\caption{\label{SM_dev1_Vg12T} (\textbf{A}) Device 1A, temperature dependence of constriction conductance at zero DC bias, $V_\text{GIL} = $ 3~V, $V_\text{BG} =$ 50~V. (\textbf{B}) Same at $B = $ 0.25~T. (\textbf{C}) Peak-to-dip crossover in DC bias characteristics as a function of magnetic field, at 364~mK, $V_\mathrm{G12}=$ 2~V. A series resistance $R_S$ = 1.15~k$\Omega$ was subtracted in (\textbf{A}-\textbf{C}).}
\end{figure*}

Fig.~\ref{SM_dev1_Vg12T} shows $V_\text{G12}$ sweeps at different temperatures for device 1A. All data shown in this section are for the same device state as in the main text: $V_\text{GIL}=$ 3~V. Unless specified otherwise, $V_\text{BG} =$ 50~V.

In Fig.~\ref{SM_dev1_Vg12T}A , the supercurrent dominates at low temperature. $T$ needs to be raised above 500~mK to access normal state conductance, with plateau-like features near $G = 2e^2/h$ and $4e^2/h$. This fits the classic pattern of a QPC with spin-degenerate ballistic modes. Fig.~\ref{SM_dev1_Vg12T}B shows a similar measurement, but in a 0.25~T field, which suppresses the supercurrent at all temperatures. The pattern of fluctuations in $G$ with a decreasing amplitude at higher $T$ is more clearly visible in this measurement. 

Fig.~\ref{SM_dev1_Vg12T}C illustrates a challenge with precisely assigning absolute value of plateau conductance in presence of the supercurrent. In the normal state at high $T$, excess conductance persists above $T_c$. Upon applying a magnetic field, a negative contribution to conductance appears above $B_c$. Above $T_c$, this manifests itself as a field-driven crossover between a peak and a dip of $G$ around $V_\text{DC}=$ 0. The same effect is responsible for traces Fig.~\ref{SM_dev1_Vg12T}A consistently lying above the ones in Fig.~\ref{SM_dev1_Vg12T}B.

\begin{figure*}
\centering
\includegraphics[width=7in]{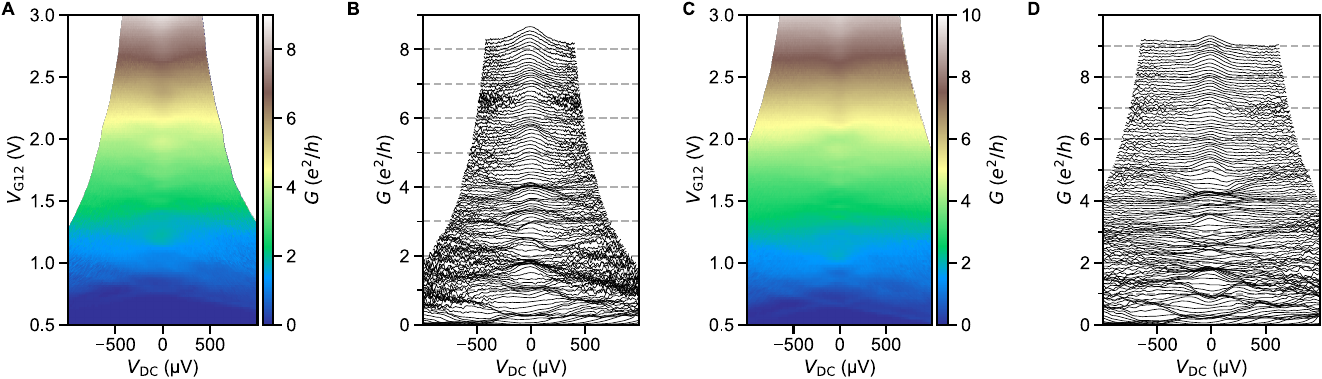}
\caption{\label{SM_dev1_Vg12Vdc_2} Constriction conductance map and corresponding line traces with $V_\text{G12}$ and DC bias. Device 1A, $V_\text{GIL} = $ 3~V, $V_\text{BG} =$ 50 V at (\textbf{A}, \textbf{B}) 511~mK, 0~T and (\textbf{C}, \textbf{D}) 364~mK, 0.1~T.  A series resistance $R_S$ = 1.15~k$\Omega$ was subtracted in (\textbf{A}-\textbf{D}).}
\end{figure*}

\begin{figure*}
\centering
\includegraphics[width=7in]{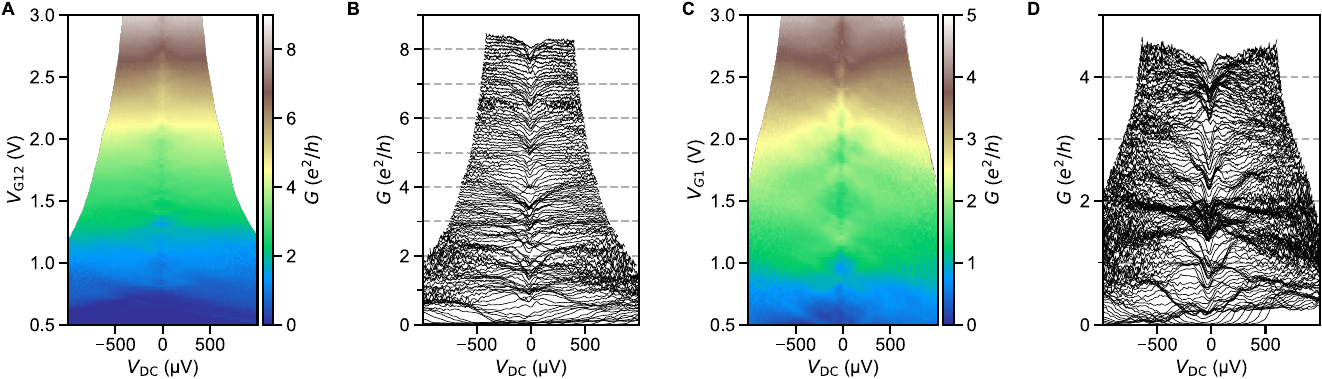}
\caption{\label{SM_dev1_Vg12Vdc_3}(\textbf{A}, \textbf{B}) Constriction conductance map and corresponding line traces with $V_\text{G12}$ and DC bias. Device 1A, $V_\text{GIL} = $ 3~V, $V_\text{BG} =$ 50~V, 45~mK, 0.5~T. (\textbf{C}, \textbf{D}) Same, but $B=$ 0.25~T and the local gate voltage is applied to only one gate ($V_\text{G1}$), the other gate ($V_\text{G2}$) is fixed at 0.9~V. A series resistance $R_S$ = 1.15~k$\Omega$ was subtracted in (\textbf{A}-\textbf{D}).}
\end{figure*}

This zero bias peak behavior persists at temperatures significantly above $T_c$. The conductance map with $V_\text{G12}$ and $V_\text{DC}$ at 511~mK in Fig.~\ref{SM_dev1_Vg12Vdc_2}A, B has zero bias peaks across entire the entire $V_\text{G12}$ range, with peak heights up to $\approx0.5 e^2/h$. Fig.~\ref{SM_dev1_Vg12Vdc_3}A and B show similarly sized dips at $B=$ 0.5 T and 45~mK. Fig.~\ref{SM_dev1_Vg12Vdc_2}C, D shows a measurement at 364~mK and 0.1~T, i.e. very close to the peak to dip crossover in Fig.~\ref{SM_dev1_Vg12T}C. It shows clear trace crowding near $G = 2e^2/h$ and $4e^2/h$ ($n = $ 1, 2), and much weaker crowding slightly below $6e^2/h$ and  $8e^2/h$  ($n = $ 3, 4).

Fig.~\ref{SM_dev1_Vg12Vdc_3}C and D show a measurement at 0.5~T, as a function of a single local gate $V_\text{G1}$, with the other gate fixed at $V_\text{G2}=$ 0.9~V. As discussed in section~\ref{sectiontunnel}, this gate trajectory bypasses a number of unintentional quantum dot resonances and thus shows a cleaner observation of trace crowding near $G = 2e^2/h$ and $4e^2/h$

\begin{figure*}
\centering
\includegraphics[width=6.8in]{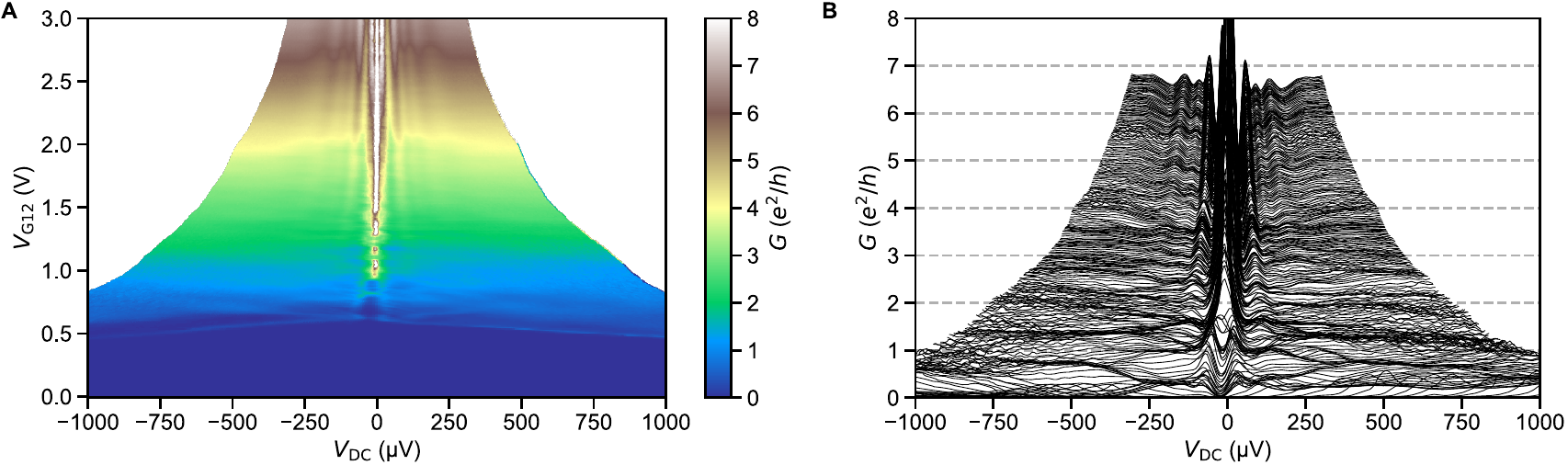}
\caption{\label{SM_dev1_Vg12Vdc_1} Constriction conductance map and corresponding line traces with $V_\text{G12}$ and DC bias. Device 1A, $V_\text{GIL} = $ 3~V, $V_\text{BG} =$ 50~V, 45~mK. No series resistance was subtracted.}
\end{figure*}

Fig.~\ref{SM_dev1_Vg12Vdc_1}A and B show the conductance map at base temperature. This measurement taken in the same state as Fig.~2 in the main text, but with smaller resolution and across a larger DC bias range. The supercurrent and the subgap structure makes plateau identification difficult in this state. Trace crowding can nevertheless be seen below $V_\text{DC} \approx $ 200 $\mu$V, near $G_N \approx$ 1.5 and 2.5 $e^2/h$ (corresponding to $n =$ 1 and 2, no $R_S$ was subtracted in this plot). Above 200 $\mu$V, most regions with trace crowding get split, as one expects for ballistic modes of a QPC \cite{Kouwenhoven89}.

\begin{figure*}
\centering
\includegraphics[width=5.3in]{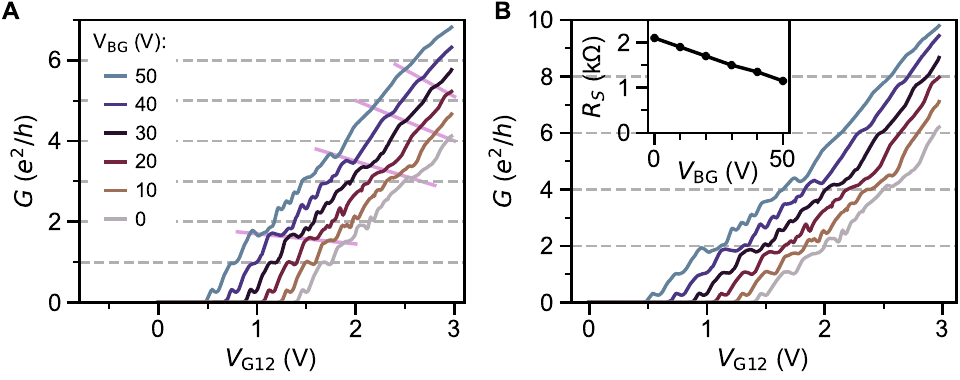}
\caption{\label{SM_dev1_Vbg} Constriction conductance traces with split gate voltage, at fixed $V_\text{BG}$= 0-50~V, taken in the normal state at T= 511~mK, $V_\text{GIL} = $ 3~V, Device 1A. (\textbf{A}) Data presented without series resistance subtraction. Solid light purple lines indicate features corresponding to $n = $ 1, 2, 3, 4. (\textbf{B}) Same data, presented with subtraction of a variable series resistance gradually decreasing from to 2.1 to 1.15~k$\Omega$ with $V_{BG}$ (shown in the inset).}
\end{figure*}

While plateau features near even multiples of $2e^2/h$ were emphasized in the above discussion of device 1A, other features can also be identified near $G = 0.2e^2/h$, $e^2/h$, and $2.5e^2/h$ (most easily seen in Fig.~\ref{SM_dev1_Vg12T} and Fig.~\ref{SM_dev1_Vbg}). As discussed in section~\ref{sectiontunnel}, these features coincide with charging resonances of an accidental Coulomb blockade. 

The overall situation also bears resemblance to the one in \cite{Mittag19}, where gate voltage tuning of the disorder potential surrounding the constriction resulted into spurious appearances of features and plateaus at odd integer multiples of $e^2/h$. In our case, the back gate voltage $V_\text{BG}$ serves a similar function by tuning the depth of the 2DES. This is seen in the data previously shown in Fig.~\ref{SM_Vbg}C. Fig.~\ref{SM_dev1_Vbg} shows the same measurement at a temperature closer to $T_c$ (511 instead of 866~mK). Fig.~\ref{SM_dev1_Vbg}B reproduces Fig.~2E in the main text. In raw data, all features in $G$ are gradually shifted downwards in $G$ as $V_{BG}$ is lowered due to a  $V_{BG}$-dependent series resistance $R_S$. Subtraction of $R_S$ that matches the features near $2e^2/h$ and $4e^2/h$ naturally aligns most other features in $G$. Plateau signatures at $n=$ 1 and 2 are present for all $V_{BG}$. At intermediate $V_\text{BG} = $ 30-40~V, faint features are also visible near $G = 6e^2/h$ and $8e^2/h$ ($n=$ 3, 4).  The Coulomb blockade features are near $G = 0.2e^2/h$, $e^2/h$, and a set of smaller fluctuations near $2.5e^2/h$. The location of the latter feature is particularly sensitive to $V_\text{BG}$, consistent with back gate tuning of the disorder potential.

\clearpage
\section{DC bias spectroscopy and critical current in the superconducting state}
\label{sectiondev123}

\begin{figure*}[b]
\centering
\includegraphics[width=6.8in]{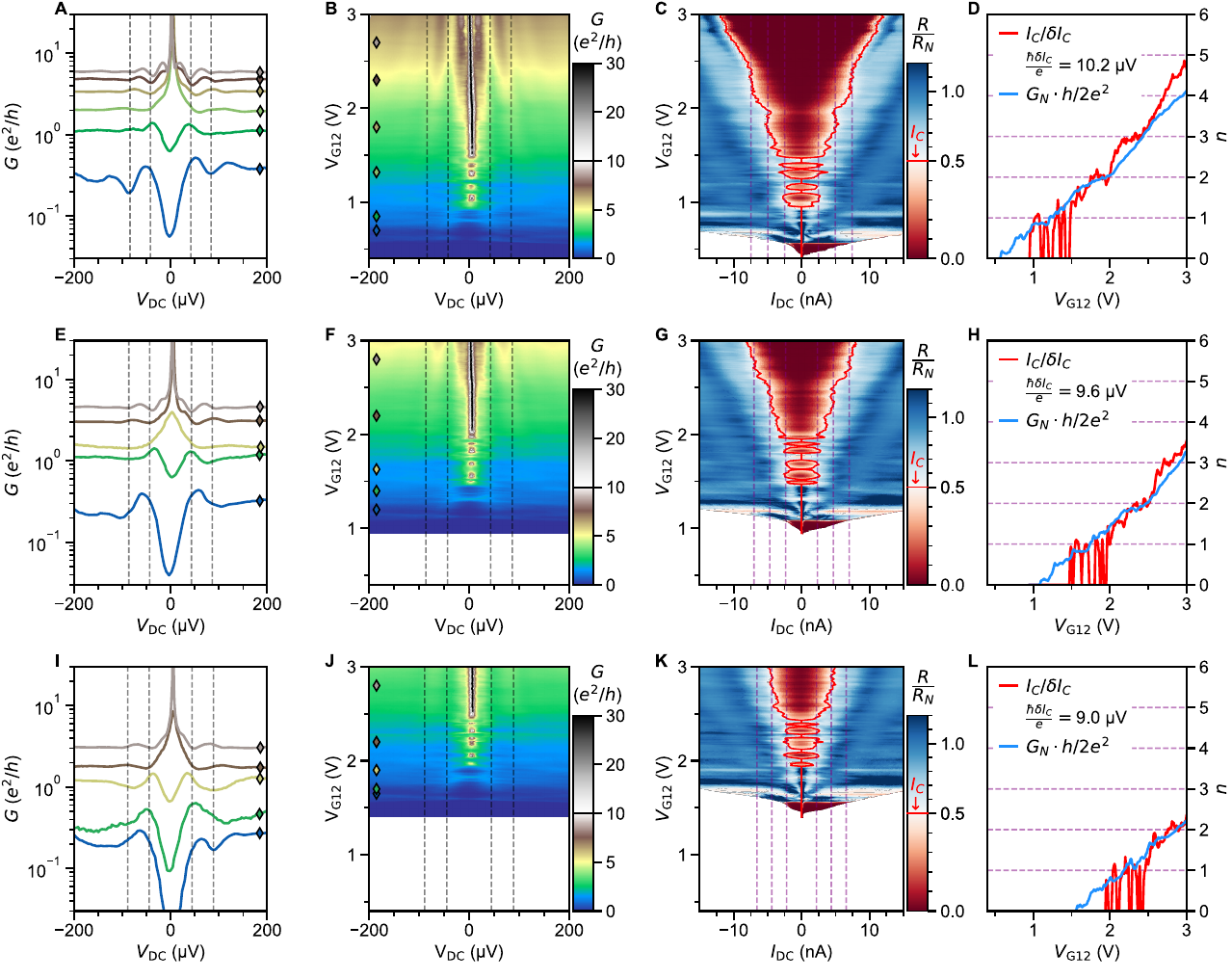}
\caption{\label{SM_vg12_vdc_1} Back gate tuning of supercurrent in device 1A. (\textbf{B}) Constriction conductance as a function of $V_\text{G12}$ and $V_\text{DC}$. Selected cuts in $V_\text{DC}$ at $V_\text{G12}$ indicated by diamond markers are plotted in (\textbf{A}). Dashed lines in (\textbf{A},\textbf{B}) indicate $\pm\Delta/e$ and $\pm 2\Delta/e$ estimated from the $T_c$ in the leads. (\textbf{C}) Constriction resistance normalized to its normal state value, taken at $V_\text{DC} = $ 100 $\mu$V. The solid red line indicates the critical current at $R=R_N/2$. The dashed lines indicate the integer multiples of the critical current quantum $\delta I_c$. (\textbf{D}) Normalized critical current and normal state conductance as a function of split gate voltage. Both quantities tracks the number of spin degenerate ballistic modes $n$ in the constriction. (\textbf{A}-\textbf{C}) is for device 1A, $V_\text{GIL}=$ 3~V, $V_\text{BG}=$ 50~V, $T$ = 45~mK. (\textbf{E}-\textbf{H}) Same as top row with $V_\text{BG}=$ 25~V. (\textbf{I}-\textbf{L}) Same as top row with $V_\text{BG}=$ 0~V. Series resistance correction is only applied to $G_N$ in the rightmost plot column: $R_S=$ 0.8~k$\Omega$ (\textbf{D}),  1.1~k$\Omega$ (\textbf{H}), 1.6~k$\Omega$ (\textbf{L}).}
\end{figure*}

This section presents conductance maps with DC bias and split gate voltage for various devices in the superconducting regime. Fig.~\ref{SM_vg12_vdc_1}A-D shows the same data as in the Fig.~2 of the main text. All of the data shown is taken in the same measurement, in which a nominal DC bias is applied to an ohmic contact, a more accurate DC bias $V_\text{DC}$ is measured at voltage probes near the constriction, and the DC current $I_\text{DC}$ is measured at the grounded ohmic contact. The selected traces of $G$ with $V_\text{DC}$ in Fig.~\ref{SM_vg12_vdc_1} illustrate the split gate-driven crossover from complete pinch-off to tunneling and Josephson junction regimes.  

\begin{figure*}[t]
\centering
\includegraphics[width=6.8in]{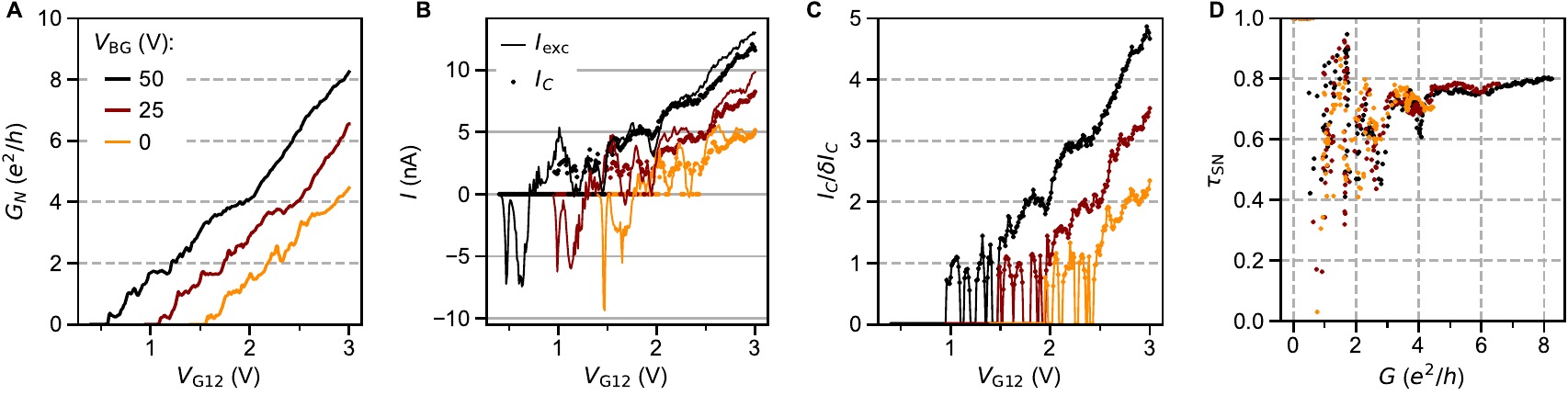}
\caption{\label{SM_vg12_vdc_1B} Direct comparison of data at $V_\text{BG} =$ 50, 25, 0~V, also shown in the different rows of Fig.~\ref{SM_vg12_vdc_1}. (\textbf{A}) Normal state conductance. (\textbf{B}) Critical current (lines) and excess current (circles). (\textbf{C}) normalized critical current. (\textbf{D}) SN contact transparency extracted from $eI_\text{exc}R_N/\Delta$, plotted as a function of normal state conductance. Correction for  $R_S$ was applied to $G_N$ in figures (\textbf{A}, \textbf{D}), with same values as in Fig.~\ref{SM_vg12_vdc_1}}
\end{figure*}

\begin{figure*}
\centering
\includegraphics[width=6.8in]{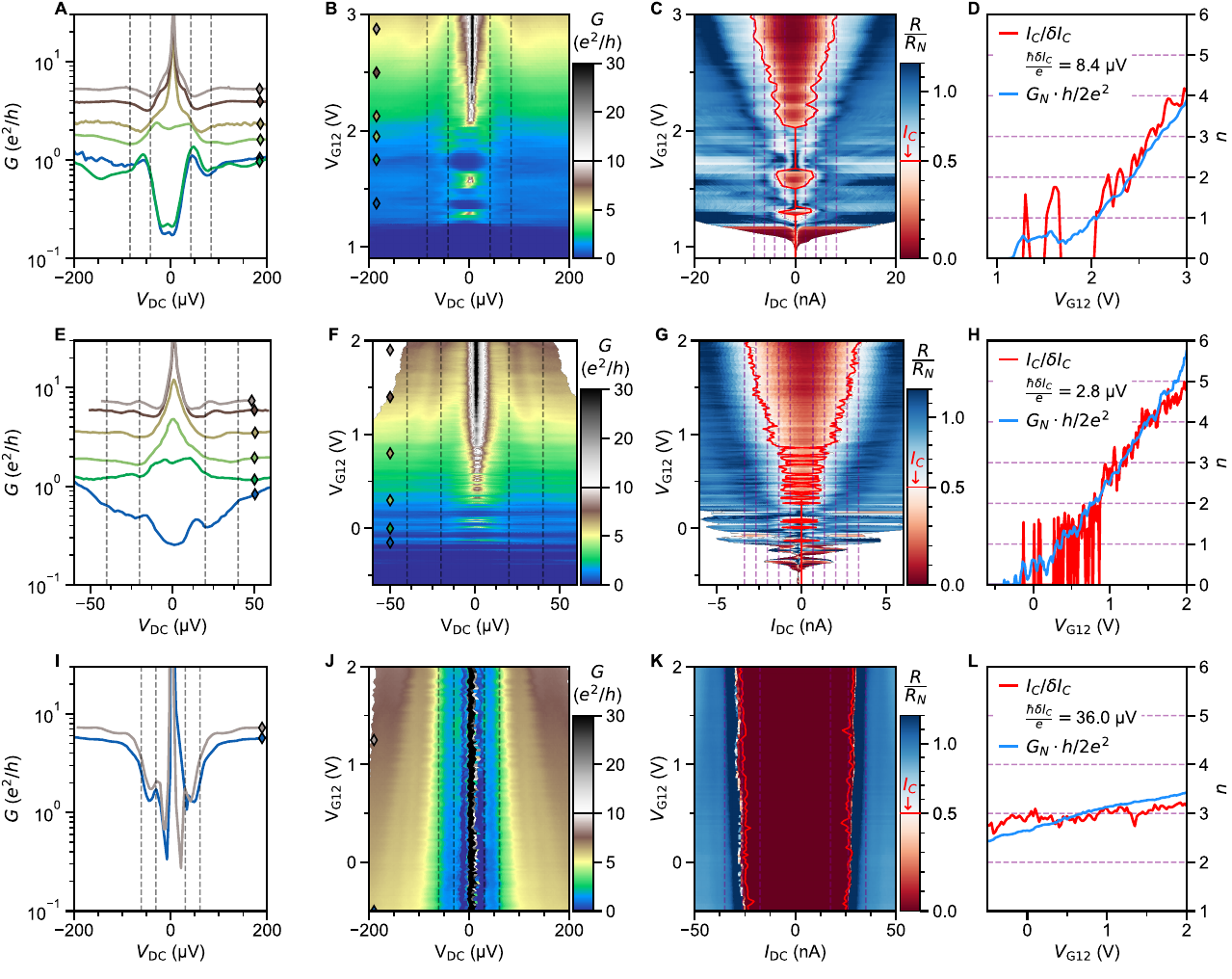}
\caption{\label{SM_vg12_vdc_2} Same data type as Fig.~\ref{SM_vg12_vdc_1}. (\textbf{A}-\textbf{D}) is for device 1B, $V_\text{GIL}=$ 3~V, $V_\text{BG}=$ 50~V, $T$ = 45~mK. (\textbf{E}-\textbf{H}) is for device 2B, $V_\text{GIL}=$ 3.5~V, $T$ = 27~mK. (\textbf{I}-\textbf{L}) is for device 2A, $V_\text{GIL}=$ 3.7~V, $T$ = 28~mK. $G_N$ is taken at 40 $\mu$V in (\textbf{G}, \textbf{H}) and 100 $\mu$V in (\textbf{C}, \textbf{D}, \textbf{K}, \textbf{L}). $G_N$ was corrected for $R_S=$ 1~k$\Omega$ in (\textbf{D}), 0.5~k$\Omega$ in (\textbf{H}), 0~k$\Omega$ in (\textbf{L}).}
\end{figure*}

\begin{figure*}
\centering
\includegraphics[width=4.5in]{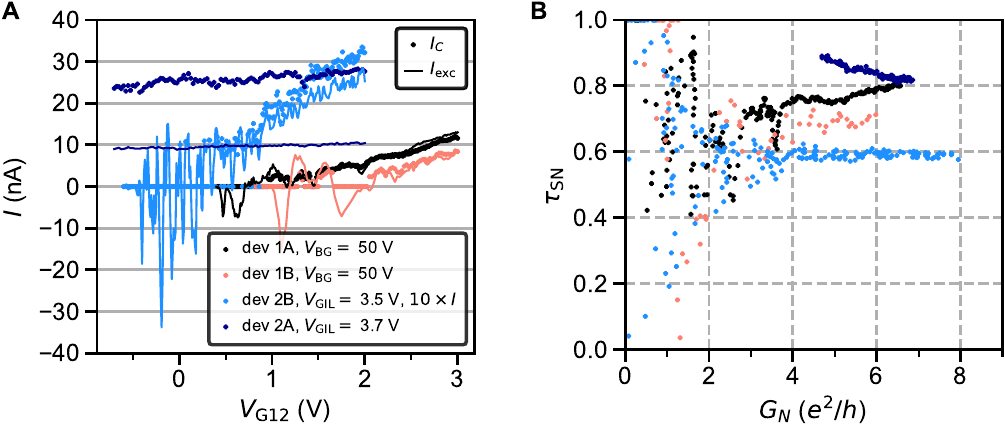}
\caption{\label{SM_vg12_vdc_2B}(\textbf{A}) Direct comparison of the critical (circles) and excess (lines) current in devices 1A, 1B, 2A, 2B. Data for device 2B is scaled by a factor of 10. (\textbf{B}) Equivalent SN transparency from $eI_\text{exc}R_N/\Delta$, with the gap value taken from $T_c$ in the leads. It is plotted as a function of normal state resistance from Fig.~\ref{SM_vg12_vdc_1} and \ref{SM_vg12_vdc_2}, without correction for $R_S$.}
\end{figure*}

\begin{figure*}
\centering
\includegraphics[width=6.8in]{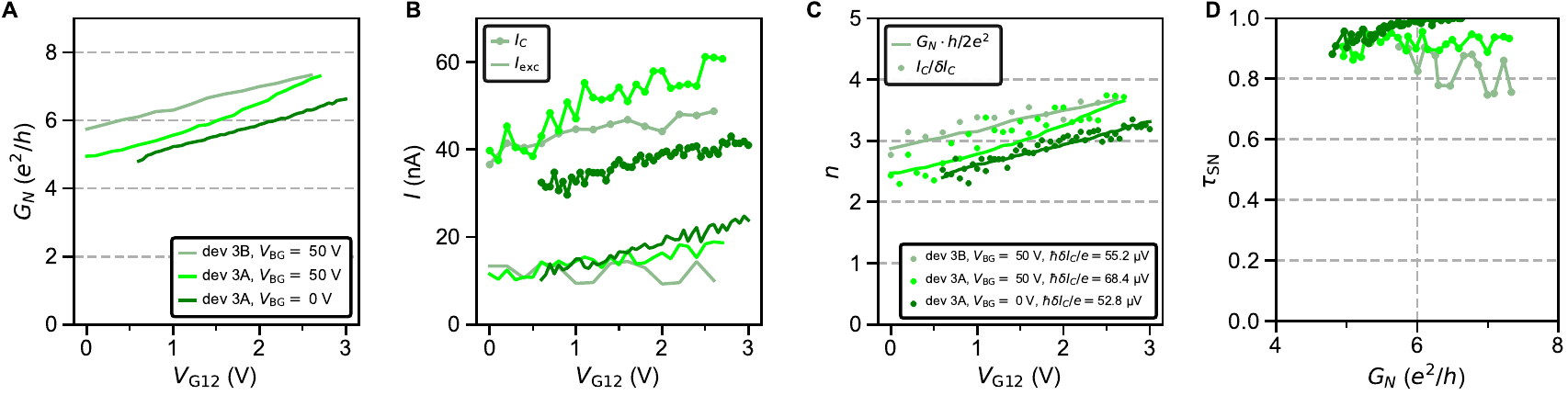}
\caption{\label{SM_vg12_vdc_3B} Devices 3A and 3B at $V_\text{GIL} =$ 3~V, $V_\text{BG} = $ 0 and 50~V, $T=$ 29~mK. (\textbf{A}) Normal state conductance at $V_\text{DC} = $ 250 $\mu$V. (\textbf{B}) Critical (connected circles) and excess (lines) current. (\textbf{C}) $I_c$ normalized by $\delta I_c$, matched to $G_N\cdot h/2e^2$. (\textbf{D}) Equivalent SN transparency from $eI_\text{exc}R_N/\Delta$, with the gap value taken from $T_c$ in the leads. No series resistance correction was applied.}
\end{figure*}

\begin{figure*}
\centering
\includegraphics[width=6.8in]{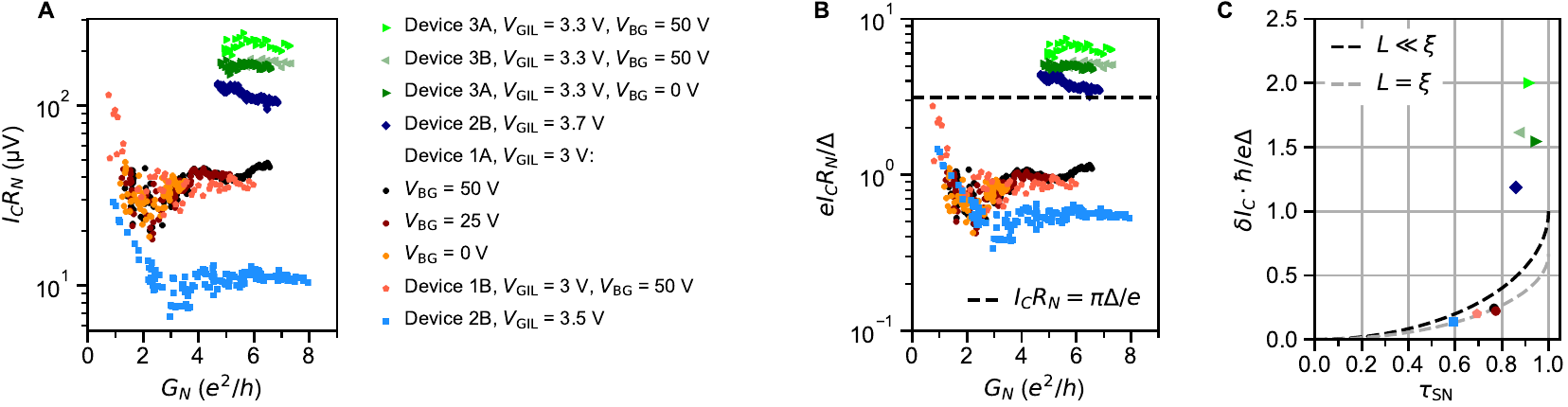}
\caption{\label{SM_vg12_vdc_3C} (\textbf{A}) $I_c R_N$ product as a function of normal state conductance. See legend for device state details. (\textbf{B}) Same as (\textbf{A}), but with $I_c R_N$ normalized to the gap value from lead $T_c$. (\textbf{C}) $\delta I_c$, the critical current per ballistic mode, plotted as a function of SN transparency (averaged in the $G_N=$ 4-6 $e^2/h$ region). Blacked dashed line is the short limit SNS model (eq.~\ref{eqSNSIc}), gray dashed line is the same model with an additional factor $\alpha = 2/3$ from finite weak link length $L=\xi$ (eq.~\ref{eqIcL2}).}
\end{figure*}

In the tunneling regime ($V_\text{G12}<$ 1~V), conductivity is suppressed near zero bias. The gap is indicated in Fig.~\ref{SM_vg12_vdc_1}A and B at $\Delta = 1.76 k_B T_c$ \cite{Swartz18}. The $T_c$ value is extracted from the measurement of lead resistance $R_\mathrm{lead}$ with temperature. Coherence peaks in conductance are seen near $V_\text{DC} = \Delta/e$. This is consistent with the expectations of tunneling across a superconductor/normal metal (SN) interface \cite{Blonder82}. In the SNS geometry of our device, this regime can be understood as tunneling across two SN interfaces in series, as discussed in \cite{Gallagher14}. The intermediate regime ($V_\text{G12} = $ 1-1.5~V) with intermittent supercurrent is discussed in section~\ref{sectiontunnel}. In the Josephson regime ($V_\text{G12} >$ 1.5~V), there is a robust supercurrent at zero bias. At $V_\text{DC} = \Delta/e$, coherence peaks at low $V_\text{G12}$ evolve into conductance dips at high $V_\mathrm{G12}$. Smaller dips can be seen near $V_\text{DC} = 2\Delta/e$. This inversion is characteristic of SNS junctions with highly transparent SN interfaces \cite{Averin95, Kjaergaard17}. 

Similarly to Fig.~2 in the main text, the analysis of the critical supercurrent is presented in Fig.~\ref{SM_vg12_vdc_1}C using constriction resistance normalized to its normal state value $R_N$, extracted in the same measurement as the resistance at $V_\text{DC} = $ 100 $\mu$V. A direct comparison between $I_c$ and $G_N$ is shown in Fig.~\ref{SM_vg12_vdc_1}D by normalizing both qunatities into a number of ballistic modes $n$. $G_N$ is divided by $\delta G_N = 2e^2/h$ under the assumption of spin degenerate ballistic modes. $I_c$ is divided by $\delta I_c = e\Delta/\hbar$, chosen to match the plateau structure in $I_c$. This plot emphasizes the numerical correspondence between these two independently measured quantities.

As shown in Fig.~\ref{SM_vg12_vdc_1}E-L, lowering the back gate voltage $V_\text{BG}$ from 50~V to 25 and 0~V shifts the constriction pinch-off point from $V_\text{G12}=$ 0.5 to 1 and 1.5~V. The patterns of tunnel to Josephson junction crossover, $I_c$ quantization, and numerical correspondence of $I_c/\delta I_c$ with $G_N$ are preserved. This is emphasized in the side-by-side comparison of key quantities at different $V_\text{BG}$, shown in Fig.~\ref{SM_vg12_vdc_1B}. $\delta I_c$ is slightly reduced from 2.48 nA ($V_\text{BG} =$ 50~V) to 2.34 (25~V) and 2.19 nA (0~V). Fig.~\ref{SM_vg12_vdc_1B}B also shows the excess current $I_\text{exc}$, which approximately follows $I_c$ for all $V_\text{BG}$. The SN interface transparency $\tau_\text{SN}$ extracted from the quantity $eI_\text{exc}R_N/\Delta$ (see section~\ref{sectionOBTK}) is shown in Fig.~\ref{SM_vg12_vdc_1B}D. When plotted as a function of constriction conductance, $\tau_\text{SN}$ overlaps for all $V_\text{BG}$, including the dips near $n =$ 1, 2, 3. These features are a natural consequence of a sharper plateau structure in $I_c$ in comparison to $G_N$.

Fig.~\ref{SM_vg12_vdc_2}A-D shows the measurement on device 1B, a slightly wider (60~nm nominal width) constriction on the same Hall bar, at $V_\text{BG} =$ 50~V.  A prolonged tunneling regime and an intermittent weak supercurrent is seen at $V_\text{G12} =$ 1-2~V. At $V_\text{G12} >$ 2V, a good correspondence numerical correspondence in $n$ from $I_c/\delta I_c$ and $G_N$ is achieved with $\delta I_c =$ 2.04 nA, a value slightly reduced but close to device 1A. The $V_\text{DC}$ dependence also shows a tunneling regime with coherence peaks near $V_\text{DC}=\Delta/e$ at low $V_\text{G12}$, that evolve into conductance dips at higher $V_\text{G12}$.

Fig.~\ref{SM_vg12_vdc_2}E-L shows data from devices 2A and 2B from separate cooldowns. Device 2B at $V_\text{GIL}=$ 3.5~V is in the strongly underdoped regime. The lead $T_c$ is 167~mK. In comparison to device 1A, The $V_\text{DC}$ dependence is re-scaled to a smaller gap, but retains the essential features: coherence peaks in the tunneling regime, conductance dips near $V_\text{DC}=\Delta/e$ and $2\Delta/e$ in the Josephson regime. The critical current in this device is also much smaller in magnitude and exhibits strong fluctuations. Its dependence on $V_\text{G12}$ is not smooth and has several short plateaus, which coincide with similar features in $G_N$. The plateau assignment is complicated by the presence of half-integer features in $G_N$, and  sensitivity to the choice of $R_S$ at high conductance. Keeping the spin-degenerate mode notation ($G = n\cdot2e^2/h$), short $I_c$ plateaus are seen at at $n =$ 3, 4 and 4.5. In the $n=$ 1, 2 region, $I_c$ is intermittent between zero and $n\cdot\delta I_c$. 

$I_c$ and $I_\text{exc}$ of this device is shown in Fig.~\ref{SM_vg12_vdc_2B}, both multiplied by 10x for comparison with devices 1A, 1B  and 2A. $I_\text{exc}$ again follows $I_c$, but lags behind it. The SN transparency extracted from $eI_\text{exc}R_N/\Delta$ is $\tau_\text{SN}$ = 0.6, appreciably lower than for devices 1A and 1B. This is consistent with the relatively low $\delta I_c$, which is suppressed by a factor of 7 in comparison to $e\Delta/\hbar$.

Device 2A at $V_\text{GIL}=$ 3.7~V is in the overdoped regime, with the lead $T_c$ at 253~mK. The gate tunability with $V_\text{G12}$ is strongly reduced in comparison to the $V_\text{GIL}=$ 3.5 V state. The constriction does not reach the pinch-off within the available range of $V_\text{G12}$. $G_N$ is only modulated between $5$ and $7e^2/h$. $I_c$ is only slightly modulated around 27 nA. With such a weak modulation, normalization by $\delta I_c$ can only be done by numerically mathching $I_c/\delta I_c$ to $G_N$. It is thus not possible to reliably establish whether the ballistic SQPC picture applies for this device. The $V_\text{DC}$ dependence of $G$ is characterized by a very large dip in conductance near $V_\text{DC} = \Delta/e$. This feature results in the excess current that is strongly decreased in comparison to $I_c$ (Fig.~\ref{SM_vg12_vdc_2B}A). The implied transparency is however very high, $\tau_\text{SN} = $ 0.8-0.9. 

A very similar situation was observed in devices 3A and 3B (Fig.~\ref{SM_vg12_vdc_3B}), which were measured in the high carrier density regime, with lead  $T_c =$ 256~mK. Both constriction conductance and $I_c$ are only weakly modulated by $V_\text{G12}$. The excess current is significantly lower than $I_c$ and the implied transparency is $\tau_\text{SN} = $ 0.8-1. In device 3A at high $V_\text{G12}$ and $V_\text{BG} =$ 0~V, $eI_\text{exc}R_N/\Delta$ exceeds 2.64, the ideal transparency limit in the SNS model.

Fig.~\ref{SM_vg12_vdc_3C} compares the $I_c R_N$ product of all devices discussed in this section. $I_c R_N$ is presented as a function of $G_N$ (as a proxy for $V_\text{G12}$) to emphasize the distinction between gate-tunable devices with lower $I_c R_N$ and weakly tunable ones with a higher $I_c R_N$. Normalizing by the superconducting gap $\Delta/e$ (from lead $T_c$) further emphasizes this clustering into two groups. This normalization is rationalized by the general expectation that $I_c R_N$ of a Josephson junction scales with the gap \cite{Golubov04}. For an ideal ballistic SNS constriction ($\tau_\text{SN}=$ 1, $L\ll\xi$), $I_c R_N = \delta I_c \cdot h/2e^2 = \pi\Delta/e$.

In experiments on SNS junctions, $I_c R_N$ is ubiquitously used as a metric for junction quality. In most casses $I_c R_N$ is substantially lower than $\Delta/e$, and $I_c R_N$ of order $\Delta/e$ is often invoked as a signature of a high quality junction 
\cite{Williams12, Mizuno13, Lee15, BenShalom16, Borzenets16, Ghatak18}.
For devices 1A and 1B, $I_c R_N$ is approximately at or slightly below $\Delta/e$, which is lower than $\pi\Delta/e$ by factor of 3-5. This statement is equivalent to the discussion in the main text on suppression of $\delta I_c$ in comparison to $e\Delta/\hbar$, provided that $G_N\cdot h/2e^2$ numerically matches with $I_c/\delta I_c$. The corresponding plot of $\delta I_c$ as a function of $\tau_\text{SN}$ is shown in Fig.~\ref{SM_vg12_vdc_3C}C. It illustrates that the data on devices 1A, 1B and 2B is consistent with the ballistic SQPC picture, with the weak link length approximately equal to or shorter than the coherence length.

For the devices in the high carrier density limit (2B at $V_\text{GIL} =$ 3.7~V, 3A, 3B), $I_c R_N$ exceeds $\pi\Delta/e$ by a factor of 1.2 - 2. A natural explanation is a crossover from an SNS junction to an SS'S constriction or wire. Establishing a crisp picture requires further study, but two frameworks can be invoked as useful starting points. On one hand, $I_c R_N$ in a superconducting wire (S') connecting two superconducting reservoirs (S) is expected to increase with length until the onset of decoherence, and can exceed $\pi\Delta/e$ \cite{Golubov04}. On the other hand, one can make a comparison to the STO leads themselves, which show ‘‘weak superconductivity’’ with a relatively small critical current. A 20x20~$\mu$m square of STO in the leads shows $eI_c R_N/\Delta = $ 4.5 near devices 1A and 1B and 3.2 near devicees 3A and 3B. In \cite{Prawiroatmodjo16}, a 50~$\mu$m long and 20~$\mu$m wide LAO/STO Hall bar has been documented to show $eI_c R_N/\Delta = $ 25-70. This was rationalized in terms of an interconnected Josephson junction array, where $I_c R_N$ scales with its size and can easily exceed $\Delta/e$. For the case of a square array of $N_\text{JJ} \times N_\text{JJ}$ junctions, $eI_c R_N/\Delta = N_\text{JJ}\pi/2$ \cite{Yu92}.

\clearpage
\section{Tunneling regime and accidental Coulomb blockade near pinch-off}
\label{sectiontunnel}
\begin{figure*}[b]
\centering
\includegraphics[width=7in]{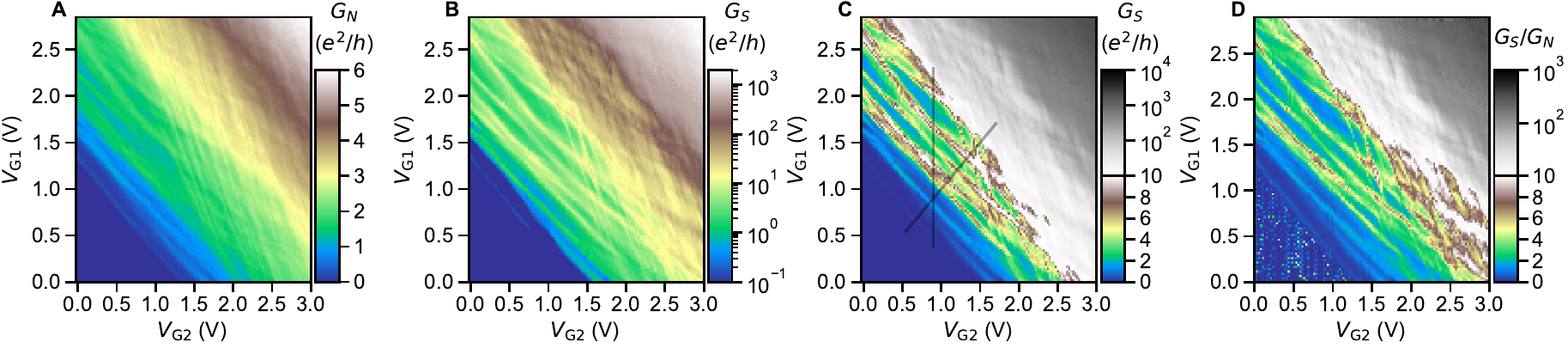}
\caption{\label{figSM_Vg1Vg2d} Potential mapping by independently sweeping the two split gate voltages $V_\text{G1}$ and $V_\text{G2}$, in device 1A at $V_\text{DC}=$  0, $T = $ 45~mK, $V_\text{GIL}=$ 3~V, $V_\text{BG}=$ 50~V. (\textbf{A}) In the normal state at $B=$ 0.25 T. (\textbf{B}) in the superconducting state at $B=$ 0. (\textbf{C}) Same as (\textbf{B}), but on a different color scale, emphasizing features at $G$\textless 10 $e^2/h$. Solid lines indicate the gate sweep trajectory in Fig.~\ref{figSM_Vg1Vdc} and \ref{figSM_Vg12Vdc} (\textbf{D}) Ratio of conductance in the superconducting and normal states.}
\end{figure*}

In the split gate geometry of our device, one can asymmetrically set the gate voltages $V_\text{G1}$ and $V_\text{G2}$. This has the effect of moving the saddle potential location around the constriction. In this manner, one can map the disorder landscape in the QPC, as shown in Fig.~\ref{figSM_Vg1Vg2d}. The conductance in both the normal ($G_N$) and superconducting states ($G_S$) rises with $V_\text{G1}$ and $V_\text{G2}$ in a largely symmetric way. This confirms that the capacitances of the two split gates of our QPC are similar, as intended.

Both $G_N$ and $G_S$ show several sets of line resonances in the $V_\text{G1}-V_\text{G2}$ space, at which conductance is increased. These resonances remain pronounced when plotting the $G_S/G_N$ ratio. In $G_S$, they are particularly pronounced at lower gate voltages and near intersections between different resonances. The intersections of these resonances correlate with the intermittent critical current seen near the first plateau ($G_N\approx 2e^2/h$) in Fig.~2 in the main text. They also coincide with the plateau-like features seen in $G_N$ (Fig.~\ref{SM_dev1_Vbg}B) near 0.2$e^2/h$, $e^2/h$  (i.e. inconsistent with the $2e^2/h$ quantization), and the smaller features near 2-2.5$e^2/h$.

\begin{figure*}
\centering
\includegraphics[width=7in]{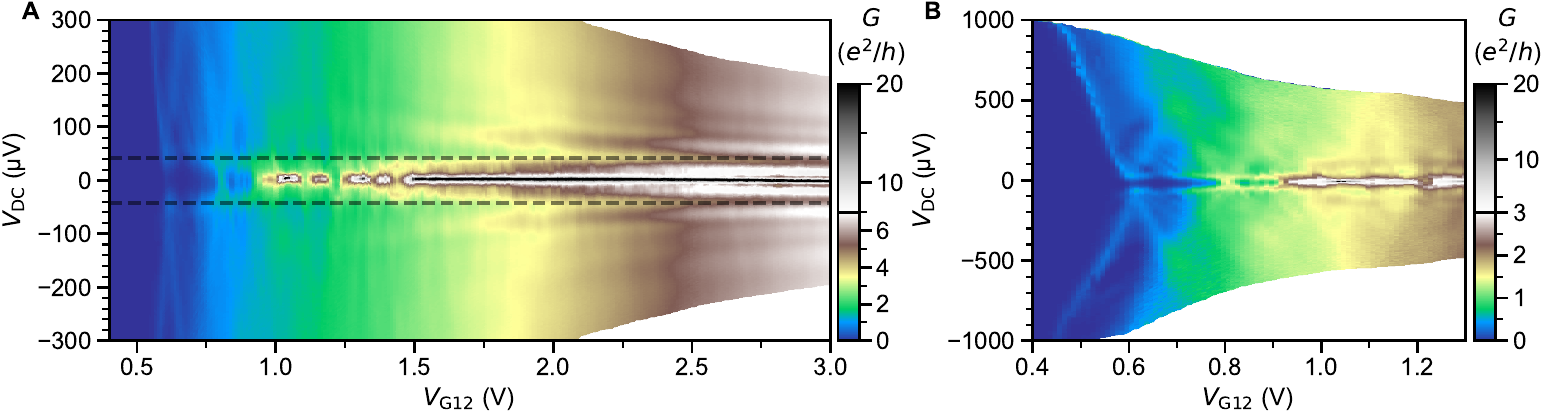}
\caption{\label{figSM_LVg12Vdc}Conductance map with DC bias and split gate voltage $V_\text{G12}=V_\text{G1}=V_\text{G2}$. (\textbf{A}) Showing the entire $V_\text{G12}$ range. The dashed lines indicate $V_\text{DC} = \pm\Delta/e$. (\textbf{B}) Same data, but focusing on the Coulomb blockade diamonds seen at low $V_\text{G12}$ and maximum $V_\text{DC}$ range. Data shown is for device 1A at $T = $ 45~mK, $V_\text{GIL}=$ 3~V, $V_\text{BG}=$ 50~V.}
\end{figure*}

We attribute these resonances to charging levels of an accidental Coulomb blockade. Spontaneous quantum dot formation near pinch-off in LAO/STO constrictions has been documented in multiple reports \cite{Maniv16,Prawiroatmodjo17,Thierschmann18}. The situation in our case is qualitatively similar. DC bias spectroscopy reveals conductance diamonds near the first two charging levels ($V_\text{G12}=$ 0.6 and 0.8 V at zero DC bias in Fig.~\ref{figSM_LVg12Vdc}B). While only 2 charging levels are clearly distinguishable before the onset of a supercurrent, their height (charging energy) starts near 400 $\mu V$ and appears to rapidly decrease with $V_\text{G12}$, following the same trend as in \cite{Cheng15,Cheng16,Maniv16,Prawiroatmodjo17,Thierschmann18}. The charging energy is likely dominated by the electrostatic capacitance of the dot rather than the orbital contribution \cite{Thierschmann18}. Its decrease with $V_\text{G12}$ can be understood as an increase in quantum dot size or tunnel barrier capacitance, although quantifying them is difficult due to the strongly electric field dependent permittivity of STO \cite{Thierschmann18}.

\begin{figure*}
\centering
\includegraphics[width=7in]{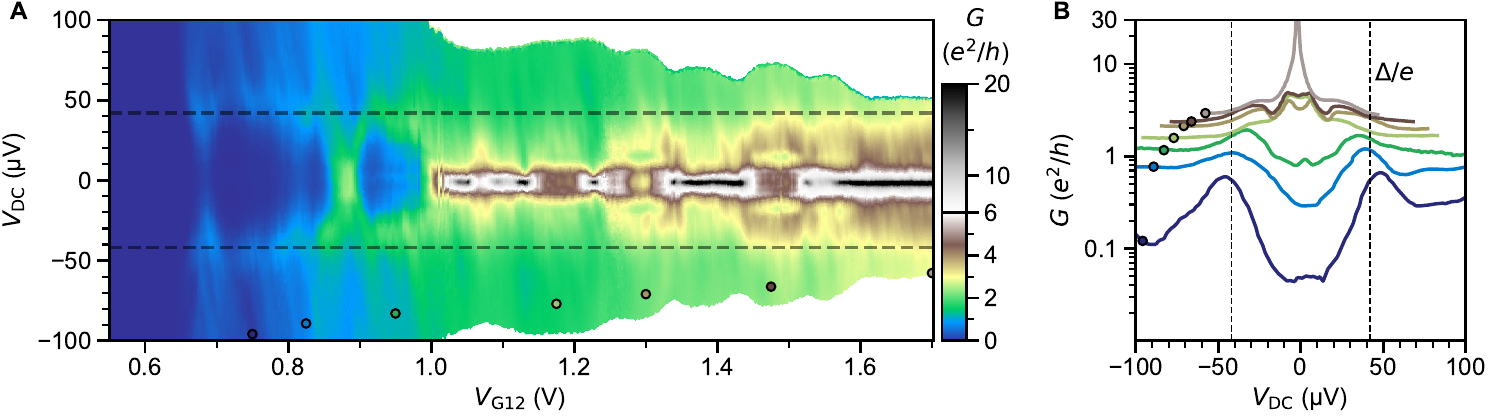}
\caption{\label{figSM_Vg12Vdc}(\textbf{A}) Conductance map with DC bias and split gate voltage $V_\text{G12}=V_\text{G1}=V_\text{G2}$, measurement range focused on the small $V_\text{DC}$ range in the tunneling and intermittent supercurrent regimes. The dashed lines indicate $V_\text{DC} = \pm\Delta/e$. Circle markers in (\textbf{A}) indicate the gate voltage position of line cuts shown in (\textbf{B}). Data shown is for device 1A at $T = $ 45~mK, $V_\text{GIL}=$ 3~V, $V_\text{BG}=$ 50~V.}
\end{figure*}

\begin{figure*}
\centering
\includegraphics[width=7in]{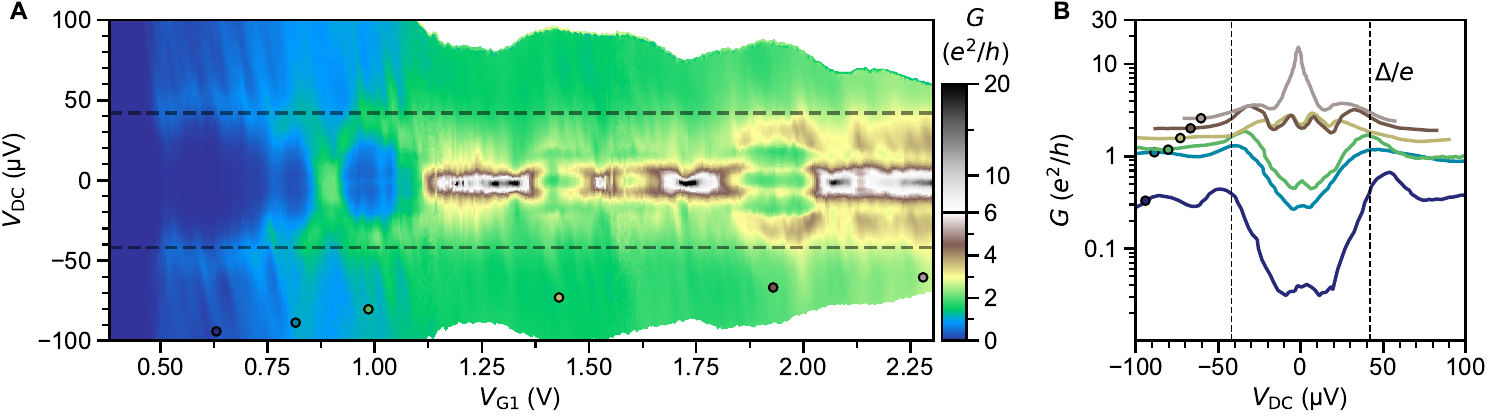}
\caption{\label{figSM_Vg1Vdc}Same as Fig.~\ref{figSM_Vg12Vdc}, but only a single split gate voltage $V_\text{G1}$ is swept, $V_\text{G2}$ is fixed at 0.9~V.}
\end{figure*}

Fig.~\ref{figSM_Vg12Vdc} shows a map with $V_\text{G12}$ and small DC bias. Fig.~\ref{figSM_Vg1Vdc} shows a similar map, but with only $V_\text{G1}$ being swept and $V_\text{G2}$ fixed at 0.9~V, a trajectory that minimizes the amount of encounters with charging resonances. In both cases, at low $V_\text{G12}$ and away from the coulomb blockade charging levels, tunneling conductance is observed: $G$ is strongly suppressed at zero bias, and coherence peaks are seen near $V_\text{DC} = \pm\Delta/e$. For comparison with experimental conductance, the gap value extracted from $T_c$ in the leads is indicated by the dashed line in Fig.~\ref{figSM_LVg12Vdc}, \ref{figSM_Vg12Vdc}, and \ref{figSM_Vg1Vdc}. This is consistent with the expectation of tunneling across an SN interface \cite{Blonder82}. Applicability to our case can be rationalized by considering the SNS junction as two SN interfaces in series \cite{Gallagher14}. The gradual increase of subgap conductance with $V_\text{G12}$ is consistent with a decrease in tunnel barrier strength \cite{Blonder82}.

Besides the peaks at $\pm\Delta/e$, tunneling conductance shows additional in-gap features: double peaks at $V_\text{DC}$ considerably lower than $\Delta/e$ ($\approx$ 7~$\mu$V), and zero bias peaks close to pinch-off. This is reminiscent of the in-gap states observed in vertical LAO/STO tunnel junctions \cite{Kuerten17}. Possible explanations involve two-band superconductivity with a small second gap \cite{Binnig80,Stornaiuolo17}, suppression of the superconducting order parameter next to the tunneling barrier due to proximity effect \cite{Golubov96,Aminov96}, Kondo effect \cite{DGG98}, Majorana or Andreev bound states \cite{Scheurer15,Kuerten17,Cheng16}. At the present stage, we do not attempt to discriminate between these possibilities.

\clearpage

\clearpage
\section{Determination of the superconducting gap}

The superconducting gap $\Delta$ in SrTiO$_3$ has been shown to be remarkably close to the BCS estimate \cite{Swartz18}. In the zero-temperature limit
\begin{equation}
\Delta_0 = 1.76 k_B T_c.
\end{equation}
The temperature dependence is well approximated by
\begin{equation}
\label{eqBCST}
\Delta(T) = \Delta_0\cdot\tanh\left(1.74\cdot\left(\frac{T_c}{T}-1\right)^{\frac{1}{2}}\right).
\end{equation}
A difficulty in our device geometry is to choose the appropriate transition temperature for estimating $\Delta$. The simplest approach is to convert the $T_c$ from the midway point of the resistance drop in the leads. For device 1A at $V_\text{BG} = $ 50~V, $T_c = $ 350~mK corresponds to $\Delta/e =$ 42 $\mu$V. This is the approach adopted throughout this paper for comparison to $\delta I_c$, $I_c R_N$, and the structure in $V_\text{DC}$ dependence of $G$.

\begin{figure*}[b]
\centering
\includegraphics[width=4in]{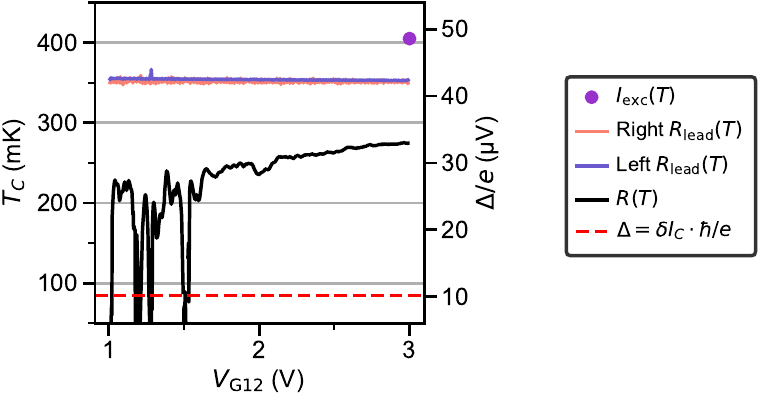}
\caption{\label{figSM_gaps} Different estimates for the superconducting gap and $T_c$ in device 1A at $V_\text{BG}=$ 50~V. Solid lines are $T_c$ from the midpoint of the resistive transition in the leads and constriction, plotted as a function of $V_\text{G12}$ (full data is shown in Fig.~\ref{SM_Vg12}). The circle marker at $V_\text{G12}=$ 3 V is the estimate from fitting the excess current to a BCS gap (Fig.~\ref{SM_VdcT2}). The dashed line is the energy scale corresponding to the $I_c$ quantization step (see Fig.~\ref{SM_vg12_vdc_1})}
\end{figure*}

This approach is based on a sheet resistance measurement, physically separated by 5 microns from the gated constriction. This sidesteps the intricacies of the electrostatic potential landscape in the immediate vicinity of the split gate. While the primary effect of $V_\text{G12}$ is to tune the carrier density in the constriction, it is likely that the electric field lines extend into the leads. Due to SrTiO$_3$ being a semiconducting superconductor, this can locally affect the $T_c$ and $\Delta$ that govern the Josephson effect of our junction. This situation is in contrast with hybrid semiconductor/superconductor systems, where the superconductor is typically a metal that is only negligibly affected by electrostatic gating.

The temperature dependence of the constriction resistance provides another estimate of $T_c$ and $\Delta$. The midpoint of the resistive transition is at 275~mK at maximum $V_\text{G12}$. It decreases to $\approx$ 240~mK near the transition to a closed constriction. This is a low estimate for $\Delta$.

\begin{figure*}
\centering
\includegraphics[width=7in]{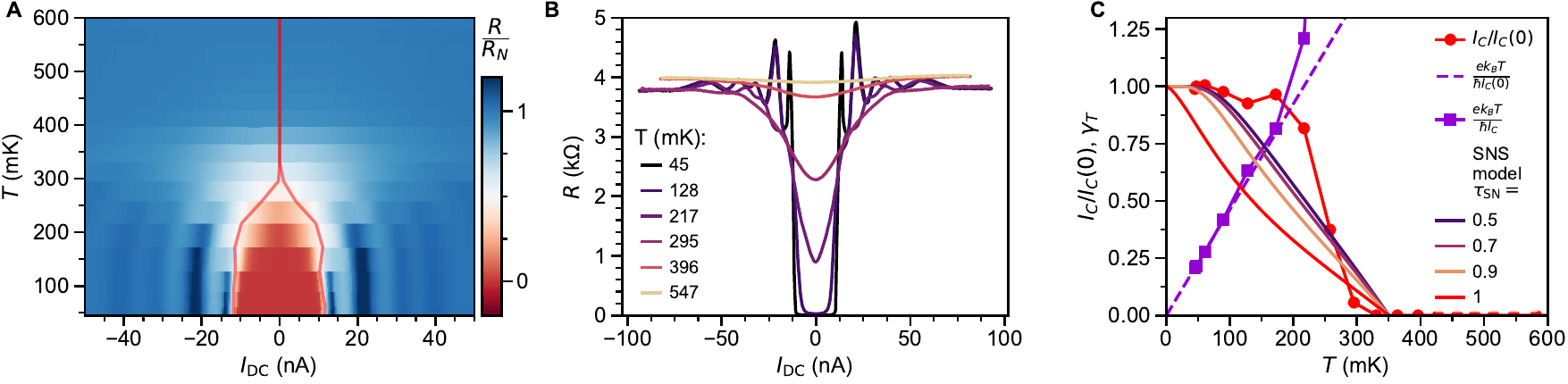}
\caption{\label{SM_VdcT1}(\textbf{A}) Temperature and DC current dependence of the constriction resistance, normalized to its normal state value at high bias. The critical current is indicated by the red solid line. (\textbf{B}) Selected cuts from the same data, plotted without normalization. (\textbf{C}) Comparison of: extracted $I_c$ normalized to the low $T$ limit $I_c(0)$ taken at 45~mK, SNS model (eq.~\ref{eqSNSIc}) with different values of SN boundary transparency, thermal broadening criterion calculated using $I_c(0)$ at base temperature and temperature-dependent values of $I_c$. All data shown are for device 1A at $V_\text{BG}=$ 50~V, $V_\text{G12}=$ 3~V.}
\end{figure*}

\begin{figure*}
\centering
\includegraphics[width=5in]{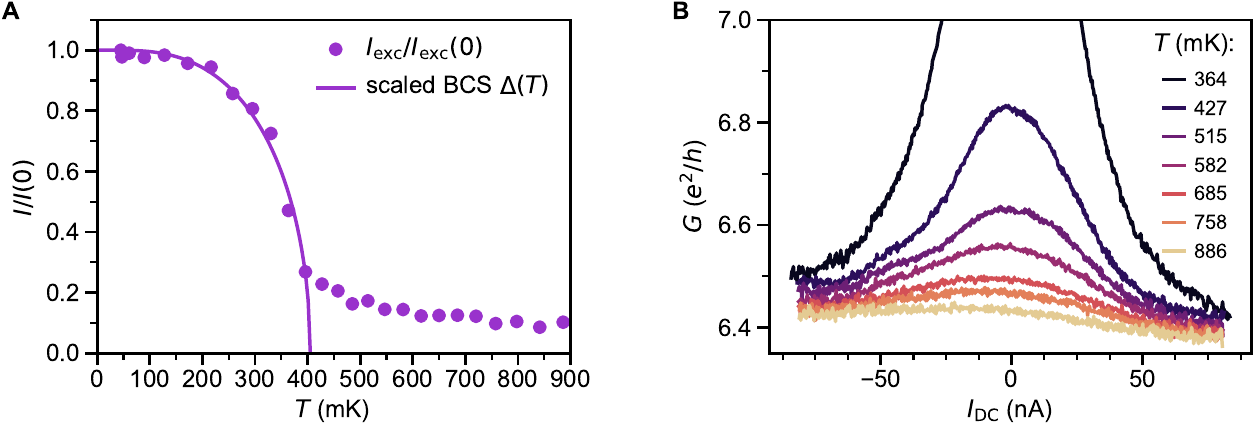}
\caption{\label{SM_VdcT2}(\textbf{A}) Temperature dependence of the excess current, normalized to base temerature value (markers). The solid line is the temperature dependence of the BCS gap with $T_c =$ 405~mK. (\textbf{B}) Selected cuts in DC current, illustrating the broadened peak in $G$ that persists above 400~mK.  All data shown are for device 1A at $V_\text{BG}=$ 50~V, $V_\text{G12}=$ 3~V.}
\end{figure*}

\begin{figure*}
\centering
\includegraphics[width=7in]{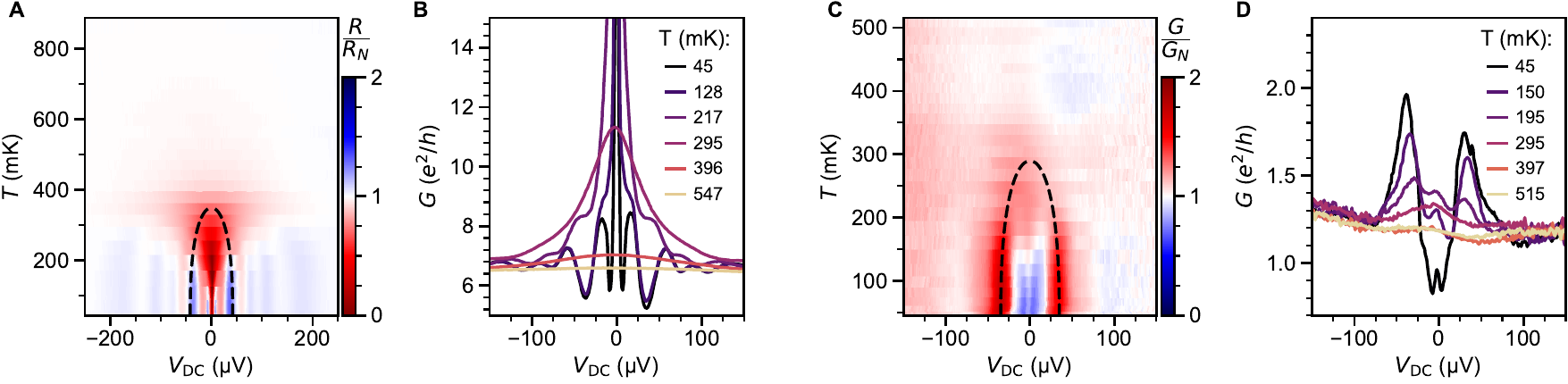}
\caption{\label{SM_VdcT3}(\textbf{A}) Device 1A, $V_\text{BG}=$ 50~V, $V_\text{G12}=$ 3 V (Josephson junction regime). Temperature and DC bias dependence of the constriction resistance, normalized to its normal state value at high temperature. The dashed line is the BCS gap dependence with $T_c=$ 350~mK, corresponding to a peak in $R$. (\textbf{B}) Selected cuts in DC bias from the same data, plotted as conductance without normalization. (\textbf{C}) Device 1A, $V_\text{BG}=$ 50~V, $V_\text{G12}=$ 0.9 V (tunneling junction regime). Temperature and DC bias dependence of the constriction conductance, normalized to its normal state value at high temperature. The dashed line is the BCS gap dependence with $T_c=$ 290~mK, corresponding to a peak in $G$. (\textbf{D}) Selected cuts in DC bias from the same data, plotted as conductance without normalization.}
\end{figure*}

A likely pitfall of this approach is thermal broadening in the supercurrent. To illustrate this, constriction resistance  as a function of temperature and DC current is shown in Fig.~\ref{SM_VdcT1}. Below 200~mK, the critical current only slightly decreases with temperature. This is consistent with the expected dependence for a short SNS (eq.~\ref{eqSNSIc}) in presence of finite SN transparency. Above 200~mK, that model does not accurately describe $I_c$. $I_c$ very briefly increases near 200~mK and quickly decreases to zero. This coincides with a rapid broadening in the $I_\text{DC}$ dependence of $G$.

In an overdamped Josephson junction ($2eI_c R_N^2C/\hbar\ll1$, with $C$ being the junction capacitance), thermal broadening is governed by the dimensionless criterion $\gamma_T=ek_BT/\hbar I_c$ \cite{Falco74}. Within this model, the supercurrent gets significantly rounded for $\gamma_T > 0.1$ and completely suppressed for $\gamma_T > 1$. As shown in Fig.~\ref{SM_VdcT1}C, $\gamma_T$ reaches 1 near 200~mK, rationalizing the rapid decrease of measured $I_c$. The overdamped regime hypothesis is consistent with a symmetric $I_c$ without any hysteresis in $I_\text{DC}$. 

Broadening alone does not explain the apparent increase in $I_c$ near $\gamma_T$, which could be a manifestation of proximity effect. In a SS'NS'S junction, where S' is a proximitized normal metal, the induced pairing gap $\Delta$' is suppressed in comparison to the bulk gap $\Delta$ in the S region. At higher temperature, the two gaps merge, increasing the relative strength of the proximity effect $\Delta$'$/\Delta$ \cite{Aminov96}.

Another way to probe the superconducting gap is to look at the temperature dependence of the excess current (Fig.~\ref{SM_VdcT1}), which is expected to scale as $I_\text{exc} ~ \Delta/eR_N$ \cite{Blonder82,Octavio83}. Below 400~mK, $I_\text{exc}$ is  well described by a scaled BCS dependence (eq.~\ref{eqBCST}) with $T_c=$ 405~mK. This provides an upper estimate for $\Delta$.

Surprisingly, $I_\text{exc}$ does not completely vanish above the $T_c$ implied by the BCS dependence. This residual $I_\text{exc}$ can also be seen as a small, heavily broadened dip in $R$ persisting above $T_c$ (Fig.~\ref{SM_VdcT2}B). We speculate that this might be a signature of pre-formed Cooper pairs without macroscopic coherence \cite{Cheng15,Dubouchet19}.

An independent confirmation of the superconducting gap can in principle be extracted from DC bias spectroscopy. Fig.~\ref{SM_VdcT3}A shows $G$ as a function of $V_\text{DC}$ and temperature. It also shows a BCS gap dependence for $T_c=$ 350~mK, which matches the $R$ peak ($G$ dip)  feature identified in Fig.~\ref{SM_vg12_vdc_1}. The temperature dependence of this feature matches the BCS prediction, but thermal broadening sets in prior to the expected decrease of $\Delta$ to zero. Consequently, from this data the transition point can only be estimated to be consistent with the 240-405~mK range discussed above. In the tunneling regime at low $V_\text{G12}$, conductance peaks corresponding to $T_c\approx$ 290~mK are clearly seen at low temperature (Fig.~\ref{SM_VdcT3}C). However, similarly to the Josephson regime at high $V_\text{G12}$, thermal broadening obscures the transition region.

In summary, the uncertainty on the gap can be summarized as follows: the middle estimate is from the resistance drop with $T$ in the leads ($T_c=$ 350~mK, $\Delta/e=$ 42 $\mu$V), the lower estimate is from the resistance drop in the constriction ($T_c\approx$ 240~mK, $\Delta/e=$ 29 $\mu$V), the high estimate is from the temperature dependence of the excess current ($T_c=$ 405~mK, $\Delta/e=$ 49 $\mu$V). Independent estimates from DC bias spectroscopy are consistent with $T_c$ falling within that range.

\end{document}